\documentclass[a4paper, 12pt]{article}
\usepackage[OT1]{fontenc}
\usepackage {amsmath}
\usepackage{amsfonts}
\usepackage{amssymb}
\usepackage{makeidx}
\usepackage{graphicx}
\usepackage{lmodern}
\usepackage{kpfonts}
\usepackage {amscd}
\usepackage {amsthm}
\usepackage {amstext}
\newtheorem {theorem} {Theorem}

\newtheorem {definition} {Definition}
\newtheorem {example}  {Example}

\newtheorem {remark} {Remark}

\numberwithin {equation}{section}

\renewcommand{\vec }[1]{\textbf {#1}}

\newcommand {\propusti} [1] {}

\begin {document}
\title {Study of internal wave breaking dependence on stratification }
\author {
    Sergey  Kshevetskii, Sergey Leble \\
    Theoretical Physics Department \\
    I.Kant Baltic Federal University \\
    email: renger@mail.ru\\
    Gdansk University of Technology, \\ Differential equations and applied mathematics department, Poland\\
email leble@mif.pg.gda.pl
}
\date {}
\maketitle

\begin{abstract}
Mixing effect in a stratified fluid is considered and examined.
Euler equations for  incompressible fluid stratified by a gravity
field are applied to state a mathematical problem and describe the effect. It is found out  that  a system of Euler
equations  is not enough  for a formulation of correct generalized
problem.   Some complementary relations are suggested  and justified.
 A numerical method  is developed and applied for study of processes of vortex
destruction and mixing progress  in a stratified fluid. The dependence of
vortex destruction on a stratification scale is investigated
numerically and it is shown that the effect increases with  the
stratification scale. It is observed that the effect of vortex destruction is
absent when  the fluid density is constant. Some simple mathematical
explanation of the effect is suggested.
\end {abstract}
\section{Introduction}

The system of Euler equations for an incompressible fluid
stratified by a gravity field is considered.  It is known that
internal gravity waves can propagate within fluids stratified by density in
a gravity field.  Internal wave breaking is one of interesting
nonlinear effects, which is characterized by  disintegration of
waves with formation  of typical spots with intensive
small-scale convection inside them.  The spots  are often  termed
convective or turbulent ones because the convection inside them
looks like   turbulence.   Static stability in the fluid becomes
recovered with the course of time, but the density  inside the
spot remains different from the surrounding stratification. The
phenomenon of internal wave breaking going with formation of a
convective spot often is termed  "internal wave mixing".

The effect  of internal wave breaking is often observed in the
ocean.  The phenomenon  is registered in the atmosphere
\cite{PfisterLSarrWCraifW} as well.
 Broken down internal waves reorganize
stratification with time. Therefore, it is impossible to
understand and explain stratification of the ocean or the
atmosphere without taking into account the effect. Intensive
intermixing of water by internal waves creates good nutrient
medium for plankton in the ocean.  Regions of the ocean, where
wave breaking happens, are generally places of heightened
biological productivity.  It stimulates growing interest to them
and to the physical effect, being cause of heightened biological
productivity.

Many papers are dedicated   to experimental study of the
phenomenon \cite{IPDDeSilvaJImbergerandGNIvey,
IPDDeSilvaHJSFernando, D.M.MasonKerswell, McEwan1971, McEwan1}. In
McEwan's experimental papers, the phenomenon  has been studied by
means of thin laser measurements.  Internal waves were excited in
a laboratory tank with sizes of $25 \, \text {cm} \times 50
\,\text {cm} \times 25 \text {cm} $ filled with a stratified
fluid.  Stratification is formed by  dependence of water
saltiness on height. The liquid density in the tank varied with
height only by 4\%. Therefore, it is possible to term this
stratification as a weak one. The composite experimental equipment
including two lasers has allowed to watch very small-scale density
fluctuations  originated
owing  to vortex destruction in the stratified fluid and to
discover many  details  of the effect under study. Observation of
small-scale structures had been based on dependence of refraction
index of light on liquid density. McEwan has discovered and
studied different stages of evolution of wave breaking and mixing
\cite{McEwan1971}, \cite{McEwan1}: overturning, development of
{interliving}
microstructure, restoring of static stability.  According to
experiments \cite {McEwan1971}, \cite{McEwan1}, the initial phase
of evolution of a nonlinear wave finishes formation of tongues of
more heavy liquid inside more light, and vice versa.  Further, the
effect of overturning develops, and tongues disintegrate into
waves of much smaller scales.  The static stability is restored in
course of time, but the fluid stratification becomes changed.

Let us discuss reasons of formation of tongues.  The velocity
field within a local vortex is inhomogeneous, and top speeds are
achieved at some distance from  the vortex center.  It leads to
formation of the tongue of a heavy liquid, permeating into strata
of a light liquid, and of the tongue of a light fluid, permeating
into strata of a heavy liquid.  Different fluid particles  cover
different path lengths for the same time, and difference of path
lengths increases with time.  If we neglect influence
of  gravity, such fluid movement  within a vortex of an elliptic
shape leads to distortion of isopycnic lines (lines of equal
density) and to reeling of isopycnic lines onto an ellipse. The
ellipse semiaxes are determined by distances from  the vortex
center up to particles having top speeds in the vortex.  Some thin
layer structure in which stratums of light and heavy fluids are
alternating is an outcome of local rotating fluid movement.
Imposed gravity field
gives birth to wave processes, and it entails breaking down
  of the regular thin layer
structure. So, we observe  mixing of the fluid.

Due to the phenomenon mechanism, whatever weak the stratification
would be, the density gradients will become arbitrary large with
the course of time.  Therefore, when we  start mathematical
modeling of the phenomenon, it is natural to presume   a solution
of nonlinear fluid equations may  be  nonsmooth and may contain
discontinuities.  It seems to be of interest to study
nondifferentiable, generalized solutions of hydrodynamic equations
of a stratified incompressible fluid placed into a gravity field.

Further, the fluid is supposed to be an ideal one.  Estimates show
that dissipative effects are important at late stages of
evolution.  Therefore, the nondissipative problem is of certain
interest.  To be sure, taking into account of dissipative terms
should  increase  solution smoothness.  The smoothing action of
dissipation is explained by inhibition  of waves with scales
smaller than some minimally possible scale dependent on
dissipative constants.  At numerical modeling, the smoothing
action of dissipation shows one's worth only if the grid step is
less than the minimally possible scale.  Often it is not
advantageous or impossible to increase resolution up to a
minimally possible scale.  It is one more argument for study of a
nondissipative problem.

Processes of mixing in viscous two-layer $3D-$ fluids with
unstable stratification were numerically explored in
\cite{ZhumayloVASinkovaOGSofronovVNStatsenkoVPandothers},
\cite{TishkinVFZmitrenkoNVLadonkinaMYePronchevaNandothers} (see
also the literature quoted there).  Authors for raising
productivity of computations have utilized parallelizing of
calculations.  The present paper essentially differs from
\cite{ZhumayloVASinkovaOGSofronovVNStatsenkoVPandothers},
\cite{TishkinVFZmitrenkoNVLadonkinaMYePronchevaNandothers} because
in mentioned papers the instability develops at the expense of
unstable stratification (a heavy liquid has been initially placed
above a light one).  In the present paper, the stratification is
continuous and stable; but in time the instability develops in itself,
that is, under other mechanism than in
\cite{ZhumayloVASinkovaOGSofronovVNStatsenkoVPandothers},
\cite{TishkinVFZmitrenkoNVLadonkinaMYePronchevaNandothers}.  The
instability is shaped by an initial vortex as it occur in occur in experiments of \cite {McEwan1971}, \cite{McEwan1}.  Furthermore, the
problem under consideration is a two-dimensional one, and
dissipative processes are not taken into account, excepting (\ref
{equationforf}). As against the mentioned papers, we are
interested not only in simulation outcomes, but also  in
substantiation of mathematical statement of the problem and in
admissible numerical methods.  It is explained by nontrivial items
in  the problem.  Researched questions are important for viscous
liquid as well.

Monographies \cite{Gabov1}, \cite{Gabov2},  and papers
\cite{Garipov1967}, \cite{GabovSveshnikov1982},
\cite{GabovMalyshevaSveshnikov1982},
\cite{KopachevskiiTemnov1984}, \cite{KopachevskiiTemnov1986},
\cite{Gabov1988} are dedicated to mathematical research of linear
equations for small-amplitude internal waves.  As for nonlinear
Euler equations for incompressible fluid, correctness of many
problems is not proved yet \cite{Zentunyan}.
Multimode approach \cite{Le}, in a finite-mode model directly related to McEwan experiment demonstrates important role of nonlinear dispersion \cite{HKL}
 We  study nonlinear
equations and  show that a system of Euler equations is not enough
for statement of a correct generalized problem in case of a
stratified fluid.  The effects of wave overturning and fluid mixing are
simulated numerically.

The study of dependence of phenomenon of vortex destructions on a
stratification scale is another  purpose of the research.  By
means of numerical experiments, we shall demonstrate the decreasing of
overturning effects with diminution of stratification scale  $H$.
On the contrary, the effect grows with increasing stratification
scale  $H$. Nevertheless, it  have  been discovered that  if we
take a limiting case when the fluid density is a constant, the
effect of vortex destruction is absent.

\section {Basic system of equations}

We suppose a fluid behavior is described by Euler equations for an
incompressible fluid
\begin{eqnarray}
    \frac {\partial \rho} {\partial t} + \frac {\partial \rho u} {\partial x} + \frac {\partial \rho w} {\partial z}  = 0, \qquad \frac {\partial \rho u} {\partial t} + \frac {\partial \rho u ^ {2}} {\partial x} + \frac {\partial \rho uw} {\partial z} =-\frac {\partial p} {\partial x}, \label {Eulerequation1} \\
    \frac {\partial \rho w} {\partial t} + \frac {\partial \rho uw} {\partial x} + \frac {\partial \rho w ^ {2}} {\partial z}  = -\frac {\partial p} {\partial z}-\rho g, \qquad \frac {\partial u} {\partial x} + \frac {\partial w} {\partial z} =0\text {.} \notag
\end{eqnarray}

Here $ \rho $ is the density, $p $ is the pressure, $u $ and $w $
are the horizontal and vertical mass velocities of the fluid, $t $
is time, $x, z $ are the horizontal and vertical coordinates
respectively,  $g $ is the free fall acceleration.

It is supposed as well that the nonperturbed fluid is stratified
in density exponentially:
\begin{equation*}
    \rho _ {0} (z) = \rho _ {00} \exp \left (-\frac {z} {H} \right), \qquad H=6.23 \,\text {m}, \qquad \rho _ {00} =1000 \,\text {kg} / {\text {m}} ^ {3}.
\end{equation*}
The chosen value of $H $ approximately corresponds to the one for McEwan's experiments \cite{McEwan1971}, \cite{McEwan1}.  In considered McEwan's experiments, the density stratification was linear, but the stratification scale was much greater than the size of the tank; therefore, the linear and exponential stratifications are practically indiscernible.  An exponential stratification is more convenient for mathematical research  of
the equations.

The boundary condition is $ \left (\vec {v}, \vec {n} \right) =0 $
at  the boundary $ \partial \Omega $ of the domain  $ \Omega $,
where $ \vec {v} = \left (u, w \right) $ and $ \vec {n} $ is a
normal line to the boundary.  Keeping in mind McEwan's
experiments, we take a rectangle with horizontal dimension $ 50 \,
\text { cm} $ and vertical dimension $ 25 \; \text { cm} $ as the
domain  $ \Omega $.

\section{Statement of the generalized problem}

\subsection {Preliminary analysis}

One of tested methods of numerical integration of Euler equations
(\ref{Eulerequation1}) was based on Galerkin method
\cite{Fletcher}.  The solution was searched for as a series in
terms of functions  dependent on vertical coordinate $z $ and
belonging to a complete orthogonal system of functions, the
expansion coefficients were dependent on horizontal coordinate $x $ and
time $t $.  From Euler equations, by known recipe we have
deduced the equations for expansion coefficients \cite{HKL}.  We  term these
equations as a set of equations for wave modes.  The set of equations
for wave modes have been solved numerically, and then the velocity
and density have been calculated using the series.

In such a way we successfully simulated an initial phase of
evolution of the nonlinear wave, which was finished with  formation
of tongue of a heavy liquid inside a light one, and vice-versa.
A typical scale of the tongues is about $2 \,\text {cm}
$.  Formation of such  tongues is (observed) watched in laboratory
experiments as well.  In laboratory experiments, these tongues
further disintegrate into waves of very small scales.  However, we
have miscarried this effect, though formally it is possible to
obtain solution of the problem for all time.

In-depth study of the problem has given the discovery that it is
impossible to simulate correctly the nonlinear process under study
within the framework of a standard Galerkin method.
 The reason is that  partial
derivatives do not exist for  functions with discontinuities.  In such cases we
should search for the nondifferentiable solution  as a generalized,
nonclassical one.
 In 1954, Lax, considering
quasilinear equations, has shown that   nonlinear equations admit
definition of different, nonequivalent classes of generalized
solutions \cite{PETERDLax1954}, \cite{Lax1957}, \cite{Richtmayer},
\cite{RichtmayerMorton}, (see also \cite{NeumannRichtmayer}).

 It is explained by the fact the main
equations are crashed  at discontinuity surfaces, so   we need
some additional conditions to match solutions on discontinuity
surfaces. However, the conditions
surface may be taken different, if we consider the question only
from  a mathematical standpoint. Therefore, convergence and
stability of the numerical method yet do not guarantee that an
obtained generalized solution is valid from the physical point of
view. Moreover, only one of possible definitions of a generalized
solution may fit the physical sense. Sufficient conditions of
physical validity of the generalized solution are not formulated
yet, but necessary requirements are obvious. A generalized solution being
adequate to the physical sense, should meet the condition of
conservation of mass, momentum and energy.

The considered equations for incompressible fluid differ from surveyed by Lax (in what sense), but Lax's argumentation is applicable to them as well.  Analysis has shown that the conventional Galerkin method does not allow achieving simultaneous conservation of mass, momentum and energy for an obtained nondifferentiable solution. May be it is possible to correct somehow conventional Galerkin method that would allow to simulate correctly solutions with discontinuities. However, the authors cannot offer acceptable adjusting of a Galerkin method. At the same time, finite-difference methods give major freedom that permits choosing a numerical scheme ensuring warranted realization of fundamental
conservation laws.

\subsubsection {Energy functionals in the theory of linearized hydrodynamic equations for the stratified fluid}

First, we  remind how the theorems of existence, uniqueness and stability for linear problems are proved. Ordinarily linear wave equations possess some quadratic functional conserved over solutions. This conserved quadratic functional is called a wave energy functional. For example, if we linearize equations (\ref{Eulerequation1}) with respect to a background density, we come to known equations of the theory of small-amplitude internal waves for which the law of wave energy conservation is  (particular in \cite{Miropolskii, Monin}):
\begin {equation}
    \frac {dH _ {lin}} {dt} =0, \qquad H _ {lin} = \frac {1} {2} \int _ {\Omega} \rho _ {0} \left (z\right) \left [\left (\, u ^ {2} +w ^ {2} \right) +g \, H \,\phi ^ {2} \right] \, \, d \,\Omega
    \label{Hlinear}
\end {equation}
Here the function $ \phi (x, z, t) $ is defined by the formula $
\rho \left (x, z, t\right) = \rho _ {0} \left (z\right) \left (1
+\phi (x, z, t) \right) $.  In (\ref {Hlinear}) the term
containing velocities means a kinetic energy, and the term
containing $ \frac {1} {2} g \, H \,\phi ^ {2} $ is a potential
energy.
 Functional (\ref {Hlinear}) is of great
importance in mathematical theory of equations for small-amplitude
waves: proofs of theorems  of existence, uniqueness, stability for
the linearized hydrodynamic equations essentially uses
conservation of functional $H _ {lin} $ over solutions.

The linearized fluid equations also possess other
conservative functional of energy
\begin {equation}
    E _ {lin} = \int _ {\Omega} \rho _ {0} \left (z\right) \left [\frac {1} {2} \left ( \, u ^ {2} +w ^ {2} \right) +g \, z \,\phi \right] \, \, d \,\Omega
    \label{linearhydrodynamicenergy1}
\end {equation}
We will name this functional a functional of hydrodynamic energy
\cite{LandauLifshits}. In (\ref{linearhydrodynamicenergy1}), the
term containing velocities means the kinetic energy and the term
containing $g \, z \,\phi $ is the potential energy.

Here both conservative energy functionals are written to underline
their distinction.  Different formulas for energy functional are
possible because energy is determined up to a constant.  If we add
 any other conservative functional to the energy functional, we
obtain a conservative functional again, which we may term an energy
functional, but the expression for energy density becomes other.
Therefore, different formulas for energy functional exist.

We have selected the form of  wave energy functional (\ref{Hlinear})  so that the energy to be zero at lack of waves.  For this
reason, functional (\ref{Hlinear}) is named wave energy
functional. Functional (\ref{linearhydrodynamicenergy1}) is
more known, but not possesses necessary mathematical properties.

\subsubsection {Wave energy functional for nonlinear equations}

Having the purpose to develop a theory for nonlinear equations, it is naturally  to extend wave energy functional (\ref{Hlinear}) to nonlinear case. The following theorem is valid.

\begin{theorem}
    Let a solution to equations (\ref {Eulerequation1}) be differentiable.  Then the functional
    \begin{equation}
        H _ {nonl} = \int\limits _ {\Omega} \left [\rho \, \frac {u ^ {2} +w ^ {2}} {2} + \rho \, g \, z + \, g \, H \,\left (\rho \, \ln \left (\frac {\rho} {\rho _ {0} \left ( 0\right)} \right) + \, \left (\rho _ {0} (z)-\rho \right) \right) \right] \, \, d \,\Omega \qquad \label{nonlimearenergy}
    \end{equation}
is conserved on solutions.  The integrand in
(\ref{nonlimearenergy}) is strictly non-negative at $ \rho \geq
0$.
The functional (\ref{nonlimearenergy}) turns into the functional
(\ref{Hlinear}), when  we take the small-amplitude limit.
\end{theorem}
\begin{proof}
Conservation of the functional (\ref{nonlimearenergy}) follows the fact that it  is the sum of conservative
functionals of the hydrodynamic energy $ \int\limits _ {\Omega}
\left (\rho \, \frac {u ^ {2} +w ^ {2}} {2} + \rho \, g \,
z\right) \, \, d \,\Omega $,  of the mass $ \int\limits _ {\Omega}
\rho \, \, d \,\Omega $, and of the conservative functionals  $
\int\limits _ {\Omega} $ $ \rho \, \ln \rho \, \, d \,\Omega $ and
$ \int\limits _ {\Omega} q (z) \, \, d \,\Omega $.
 Nonnegativity
of the term  $ \rho \, \frac {u ^ {2} +w ^ {2}} {2} $  is obvious
at $ \rho > 0 $.
To  prove  nonnegativity of remaining summands of the
integrand, we consider
    \begin{equation*}
        a = \rho \, g \, z + \, g \, H \,\left (\rho \, \ln \left (\frac {\rho} {\rho _ {0} \left ( 0\right)} \right) + \, \left (\rho _ {0} (z)-\rho \right) \right)
    \end{equation*}
    We substitute $ \rho = \rho _ {0} (z) \, \left (1 +\eta \right) $. To be sure, the requirement $ \rho> 0 $ is equivalent to $ \eta>-1 $.  For derivative with respect to variable $ \eta $, it is valid:
    \begin{equation*}
        \frac {\partial a} {\partial \eta} =H \, ln (1 +\eta) \, \rho _ {0} (z) \, g, \qquad \frac {\partial ^ {2} a} {\partial \eta ^{2}} =H \,\frac {1} {(1 +\eta) \,} \, \rho _ {0} (z) \, g \; > 0 \qquad \text{ at } \eta >-1.
    \end{equation*}
    Unique minimum of the function $a(\eta) $ corresponds to $ \eta =0 $.  It is enough to be convinced that in the limit of small-amplitude waves the functional under study turns into (\ref{Hlinear}).
\end {proof}

We term  the functional (\ref{nonlimearenergy}) as a generalized functional of
wave energy because in a small-amplitude limit it turns into wave
energy functional (\ref{Hlinear}) for linearized equations.  At
lack of wave disturbances, the integrand in \ref{nonlimearenergy})
is equal zero.  The idea of functional (\ref{nonlimearenergy})
  form, is taken from the Hamilton formalism explained in
\cite{Miropolskii}.  Nevertheless, nonnegativity of the integrand,
also some other important mathematical properties of the
functional are discovered anew.

Let us generalize (\ref{nonlimearenergy}) to a case when a solution may be nondifferentiable.  The functionals of mass and hydrodynamic energy from which the functional is constructed (\ref{nonlimearenergy}), should be conserved   for nondifferentiable solutions; it follows from their physical sense. However the functional \\ $ \int\limits _ {\Omega} \left [\, g \, H \,\left (\rho \, \ln \left (\frac {\rho} {\rho _ {0} \left (0\right)} \right) \right) \right] \, \, d \,\Omega $ is not conserved in a case if the density field is nondifferentiable. Therefore, we demand that the        following inequality have to be fulfilled
\begin{equation}
    \frac {dH _ {nonl}} {dt} \leq 0. \label{energydissipation}
\end{equation}

The sign "$ \leq $"  is chosen because  the sign "$> $"  in (\ref{energydissipation}) leads to an unstable problem.  We will use    (\ref{nonlimearenergy}) and  (\ref{energydissipation}) as a basis of the theory of generalized solutions to nonlinear equations.   Inapplicability of Hamilton formalism for considering of solutions with discontinuities of density follows from (\ref{nonlimearenergy}),  (\ref{energydissipation}).  

\subsubsection {Generalized solutions to the nonlinear problem}

\paragraph {Preliminary analysis.}

The equations (\ref {Eulerequation1}) are written in a conservative form, in the form of fundamental conservation laws. From the physical point of view, the set of equations (\ref{Eulerequation1}) is incomplete because it does not include the energy conservation law.  Let $ \omega \subset \Omega $ be an arbitrary star domain with piecewise smooth boundary, and let  $S( \omega) $ be a boundary surface of the domain  $ \omega$, $  d \vec {S} =dS \,\vec {n} $, where $dS $ is a surface element of $S (\omega ) $, and $ \vec {n} $ is an outer  normal vector  to  $S (\omega) $.  We write an integral relation for  energy
\begin {equation}
 \int\limits _ {\omega} e (x, z, t) d\Omega-\int\limits _ {\omega} e (x, z, 0) d\Omega -\int\limits _ {0} ^ {t} \oint\limits _ {S\left (\omega \right)} e (x, z, t) \, \vec {v} (x, z, t) \, d\vec {S} \, dt=0, \quad e = \rho \left (\frac {u ^ {2} +w ^ {2}} {2} +gz\right) \label{secondenergydefinition}
 \end {equation}
 If the solution of our problem is differentiable, then (\ref{secondenergydefinition}) follows from (\ref{Eulerequation1}); but if the solution does not belong to a class of differentiable functions, then relation (\ref{secondenergydefinition}) does not follow (\ref{Eulerequation1}).  Therefore, permitting nondifferentiable solutions, we have to include (\ref{secondenergydefinition}) into
the system of general  equations as an individual equation.

Eqations (\ref{Eulerequation1}), (\ref{secondenergydefinition}) are not enough for statement of a correct problem.  For stability of the nonlinear problem, relation (\ref{energydissipation}) should be fulfilled as well.  As conservation of energy functional (\ref{secondenergydefinition}) is postulated,  the inequality
\begin {equation}
    \frac {\partial} {\partial t} \left (f\left (\rho \right) \right) + \nabla \left (f\left (\rho \right) \vec {v} \right) = \nu \leq 0 , \qquad f\left (\rho \right) = \rho \ln \left (\rho \right),  \label{equationforf}
\end {equation}
follows (\ref{energydissipation}) and (\ref{secondenergydefinition}).
Apparent requirement of compatibility of the relation (\ref{equationforf}) with the continuity equation  (\ref{Eulerequation1}) results in conclusion that $ \nu $ may be nonzero only at the points where $\rho$ is  nondifferentiable.

 Further we shall use  $f\left (\rho \right) = \rho \ln \left
(\frac {\rho} {\rho_0 (0)} \right) $, where $\rho_0 (0)$ is the maximum value of the density (on the bottom),

  use  of the function $f $
of fixed sign is more convenient.

\paragraph {Definition of a generalized solution.}
\label{defin}

Using the set of equations (\ref{Eulerequation1}), we  routinely
build some integral relations for definition of a generalized
solution
\begin{gather}
\int\limits _ {\Omega} \rho _ {s} (x, z, 0) \rho (x, z, 0) \,
d\Omega + \int\limits _ {Q _ {T}} \left (\rho \frac {\partial \rho
_ {s}} {\partial t} + \rho u\frac {\partial \rho _ {s}} {\partial
x} + \rho w\frac {\partial \rho _ {s}} {\partial z} \right) \, dQ
_ {T} =0, \propusti {\tag{firstdefinition}}
\label{firstdefinition} \\
\int\limits _ {\Omega} u (x, z, 0) \rho (x, z, 0) \, u _ {s} (x,
z, 0) \, d\Omega + \int\limits _ {\Omega} w (x, z, 0) \rho (x, z,
0) w _ {s} (x, z, 0) \, \, d\Omega + \notag
\\
\int\limits _ {Q _ {T}} \left [\rho u \,\frac {\partial u _ {s}
(x, z, t)} {\partial t} + \rho u ^ {2} \, \frac {\partial u _ {s}
(x, z, t)} {\partial x} + \rho uw \,\frac {\partial u _ {s} (x, z,
t)} {\partial
z} \right] \, \, dQ _ {T} \notag \\
+ \int\limits _ {Q _ {T}} \left [\rho w\frac {\partial w _ {s} (x,
z, t)} {\partial t} + \rho uw\frac {\partial w _ {s} (x, z, t)}
{\partial x} + \rho w ^ {2} \frac {\partial
w _ {s} (x, z, t)} {\partial z} + \rho gw _ {s} (x, z, t) \right] dQ _ {T} =0.\propusti {\\
u = \frac {\partial \psi} {\partial z}, \qquad w =-\frac {\partial
\\} {\partial x}, \notag} \notag
\end{gather}
The relations should be fulfilled for any $ \rho _ {s} (x, z, t)
\in C _ {0} ^ {\infty} \left (\Omega \right) \times C ^ {\infty}
[0, T] $, $u _ {s} (x, z, t) = \frac {\partial \psi _ {s}}
{\partial z} $, $w _ {s} (x, z, t) =-\frac {\partial \psi _ {s}}
{\partial x} $, $ \psi _ {s} (x, z, t) \in C _ {0} ^ {\infty}
\left (\Omega \right) \times C ^ {\infty} [0, T] $, $ \rho _ {s}
(x, z, T) = \psi _ {s} (x, z, T) =0 $.  In  equations (\ref
{firstdefinition}),  $Q _ {T} = \Omega \times \left [0, T\right]
$. For the basic functions (velocity components and density), it
  is convenient to introduce the stream function,
$u (x, z, t) = \frac {\partial \psi (x, z, t)} {\partial z} $, $w
(x, z, t) =-\frac {\partial \psi (x, z, t)} {\partial x} $, where
$ \psi $  is a curve in $ \mathring {W} _ {2} ^ {1} (\Omega) $
parameterized with time $t $ \cite{Richtmayer}, \cite{Sobolev},
\cite{Ladyzenskaya-1973}.

However, the  above integral  relations are not sufficient for
unambiguous definition of a generalized solution.  To be sure, we
should supply the  equations (\ref{firstdefinition})  with the  relation
(\ref{secondenergydefinition}) and the equation (\ref{equationforf})
in an integral form:
\begin{equation}
\int\limits _ {\omega} f\left (\rho \left (x, z, t\right) \right)
\, d\Omega -\int\limits _ {\omega} f\left (\rho \left (x, z,
0\right) \right) \, d\Omega + \int\limits _ {0} ^ {t} \oint\limits
_ {S (\omega)} f\left (\rho \left ( x, z, t\right) \, \right) \vec
{v} \left (x, z, t\right) \, d\vec {S} \, dt =\int\limits _ {0} ^
{t} \int\limits _ {\omega } \nu \left (x, z, t\right) d\Omega
dt\leq 0.\qquad \propusti {\tag{integralequationforf}}
\label{integralequationforf}
\end{equation}
Here $S (\omega) $ is a surface of star domain $ \omega $ with
smooth boundary.  Below, we  formulate working definition of the
generalized solution.

We construct here a definition of a generalized solution, which
should be acceptable from physical point of view and mathematically correct.
  From the previous reviewing, it is
evident that (\ref {firstdefinition}) is not enough to find
uniquely a physically justified generalized solution.  It is
necessary to supply  (\ref {firstdefinition}) with requirements
(\ref {secondenergydefinition}), (\ref {integralequationforf}), or
with equivalent ones.  If
\begin {eqnarray}
\rho (x, z, t), \, \rho (x, z, t) u ^ {2} (x, z, t), \, \rho (x,
z, t) w ^ {2} (x, z, t) \, \in
L _ {1} \left (Q _ {T} \right), \label {classfuncciy} \\
\rho (x, z, 0), \, \rho (x, z, 0) u ^ {2} (x, z, 0), \, \rho (x,
z, 0) w ^ {2} (x, z, 0) \, \in L _ {1} \left (\Omega \right),
\notag
\end {eqnarray}
then relations (\ref{firstdefinition}) have clear mathematical
sense. Requirements (\ref{classfuncciy}) seem to be natural.
Therefore, at analysis of  (\ref{secondenergydefinition}), (\ref{integralequationforf}),
we assume that requirements (\ref{classfuncciy}) are fulfilled.  If the solution is
piecewise-continuous and restricted in $Q_T $, then all integrals
in (\ref{integralequationforf}), (\ref{secondenergydefinition})
exist; and  it follows (\ref{integralequationforf}), that
(\ref{secondenergydefinition})
\begin{eqnarray}
\int _ {\Omega} e\left (x, z, t\right) \, d {\Omega}  =
 \int _
{\Omega} e\left (
x, z, 0\right) \, d {\Omega}. \label{conditiontoe} \\
\int\limits _ {\Omega} f (x, z, t _ {1}) d\Omega  \leq \int\limits
_ {\Omega } f (x, z, t _ {2}) d\Omega \qquad   \text{for \; all}
\quad t _ {1}> t _ {2}, \label{conditiontof}
\end{eqnarray}
A hypothetical case when requirements (\ref{classfuncciy}) are
fulfilled but the solution belongs to a class of functions, for
which the surface integral in (\ref{secondenergydefinition}) is
not defined, deserves special examination.  In this case, it is
possible to try to interpret a surface integral in
(\ref{secondenergydefinition}) as a pair of functionals ( $x $ and
$z $ components).  It is important that these functionals exist
for any star-shaped  domain $ \omega $ with smooth boundary $S (\omega) $
and the functionals are additive ones.  That is, if $ \omega = \omega _ {1} \cup
\omega _ {2} $, $ \omega _ {1} \cap \omega _ {2} = \emptyset $,
then a functional over surface $S (\omega) $ is equal to
sum of functionals over surfaces $S (\omega _ {1}) $ and $S
(\omega _ {2}) $, and relations (\ref {conditiontoe}) and
(\ref{secondenergydefinition}) are  fulfilled.  One can
easily be convinced that if requirements (\ref {classfuncciy}) and
(\ref{conditiontoe}) are fulfilled, then   such functionals can
always be defined.  Therefore, for definition of the generalized
solution, we demand realization of (\ref{classfuncciy}),
(\ref{conditiontoe}); at the same time, relation
(\ref{secondenergydefinition}) is not utilized in explicit form.
Analogous analysis reveals that condition (\ref{conditiontof}) has
to play an important role in definition of the generalized
solution, but it is possible to not use   (\ref{integralequationforf}) resulted in (\ref{conditiontof}).

\begin{definition}
\label{definit1} Let $ \vec {v} (x, z, t) = \left (u (x, z, t), w
(x, z, t) \right) $. We  term $ \lambda (x, z, t) = \left (\begin
{array} {c} \rho (x, z, t) \\ \vec {v} (x, z, t) \end {array}
\right) $ as a generalized solution of  equations (\ref{Eulerequation1}), if $u (x, z, t) = \frac {\partial \psi (x, z,
t)} {\partial z} $, $w (x, z, t) =-\frac {\partial \psi (x, z, t)}
{\partial x} $, $ \left. \psi \right\vert _ {\partial \Omega} =0 $
and if (\ref{firstdefinition}) (\ref{classfuncciy}),   (\ref{conditiontoe}),
(\ref{conditiontof}),   are fulfilled for
any
$ \rho _ {s} (x, z, t) \in C _ {0} ^ {\infty} \left (\Omega
\right) \times C ^ {\infty} [0, T] $, $u _ {s} (x, z, t) = \frac
{\partial \psi _ {s}} {\partial z} $, $w _ {s} (x, z, t) =-\frac
{\partial \psi _ {s}} {\partial x} $, $ \psi _ {s} (x, z, t) \in C
_ {0} ^ {\infty} \left (\Omega \right) \times C ^ {\infty} [0, T]$,
 $ \rho _ {s} (x, z, T) = \psi _ {s} (x, z, T) =0 $.
\end{definition}

\begin{remark}
For Navier-Stokes equations, the energy
conservation law is not included into the  system of constitutive
equations, but the solution  belongs  the class  $
\mathring {V} _ {2} \left (Q _ {T}  \right) $
\cite{Ladyzenskaya-1970};  and the energy relation follows
Navier-Stokes equations  for functions belonging to  $ \mathring {V} _ {2}
\left (Q _ {T} \right) $ \cite{Ladyzenskaya-1970}, \cite{Temam}.
In our case, an energy relation does not follow from integral
relations (\ref{firstdefinition}), and the energy relation is
imposed independently.  Per se, we apply the condition of
conservativeness (conservation of mass, energy) widely used  in
gas dynamics. In fact only the condition  (\ref{conditiontof})
important for inhomogeneous liquid, is novel.  The condition
(\ref{conditiontof}) is important  for viscous liquid as well.
Fluid of constant density $\rho \equiv 1$ was studied in known
monographs \cite{Ladyzenskaya-1970}, \cite{Temam}.
\end{remark}

\section { Simulation of vortex   destruction}

We take the initial condition:
\begin{equation}
\rho \left( x,z,0\right) = \rho_{0} \left( z\right),\qquad \psi
\left( x,z,0\right) =A \, \exp \left\{ -\left[ \left(
\frac{x-x_{0}}{l_{x}}\right) ^{2}+\left(
\frac{z-z_{0}}{l_{z}}\right) ^{2}\right]\right\} ,\qquad
\label{initialconditions}
\end{equation}
where $l _ {x} =0.16 \; \text {m} $, $l _ {z} =0.052 \;\text {m}
$, $A =-0.0095 \; \text {m} ^ {2}/\text {s} $, $x _ {0} =0.5 \;
\text {m} $, $z _ {0} =0.125 \; \text {m} $.  This initial
condition approximately corresponds to the disturbance created
from an oscillating paddle in the tank in MackEwan's experiments \cite{McEwan1971},
\cite{McEwan1}. The horizontal scale is
approximately twice more than the vertical one.  The amplitude of
a vertical velocity is of $5 \;\text {cm/s} $, the amplitude of a
level velocity is of $16 \;\text {cm/s} $.  The top speed of internal wave
propagation in the tank within a linear approximation is $c =10
\;\text {cm/s} $.  That is, the fluid velocity amplitude exceeds
the limit of linear wave propagation speed.

\begin{figure}[htbp]
\includegraphics [width=12.1166cm] {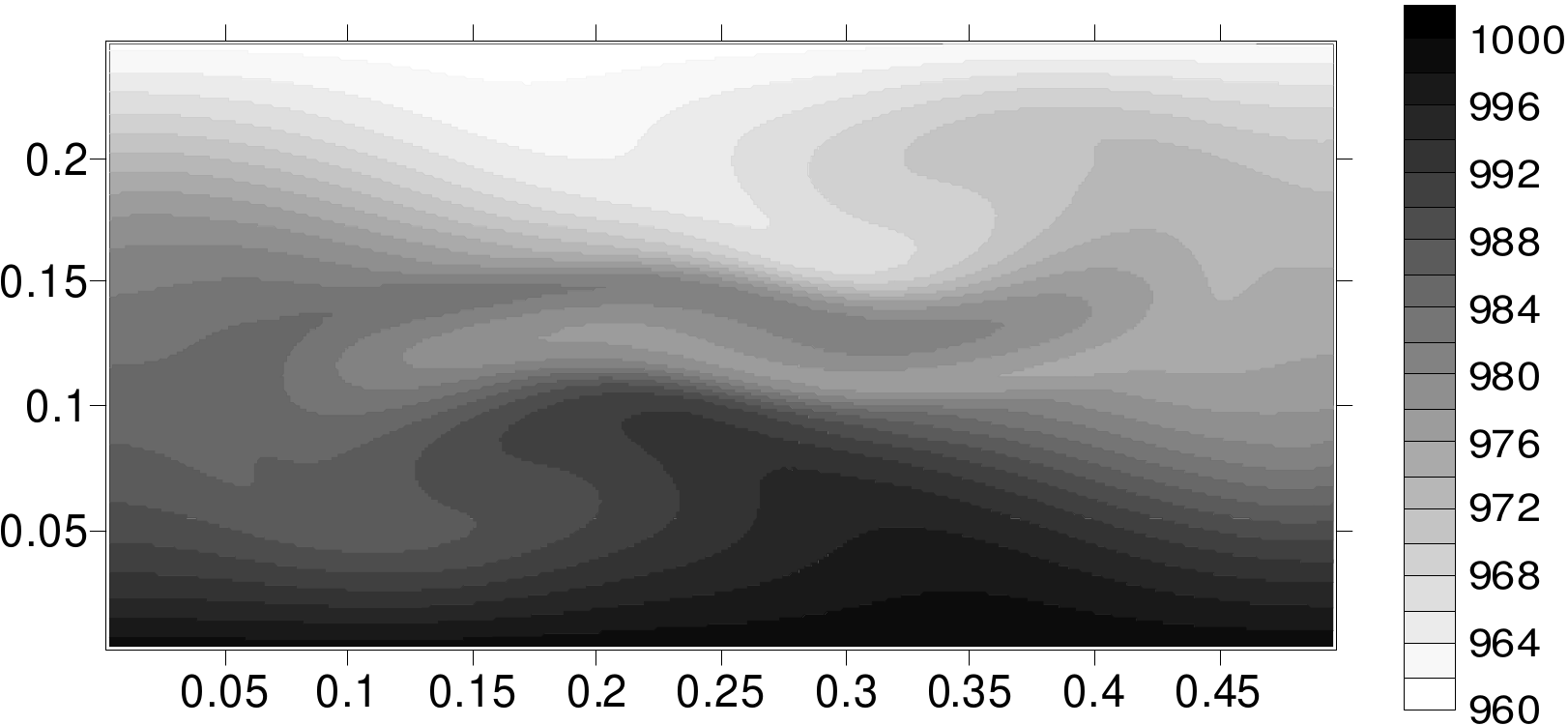}
\caption {Fluid density  at $t=3 \, \text {s} $. $H=6.23 \, \text
{m} $,  $h=0.25 \, \text {cm} $.} \label{r1-18}
\end{figure}

\begin{figure} [htbp]
\includegraphics [width=12.1166cm] {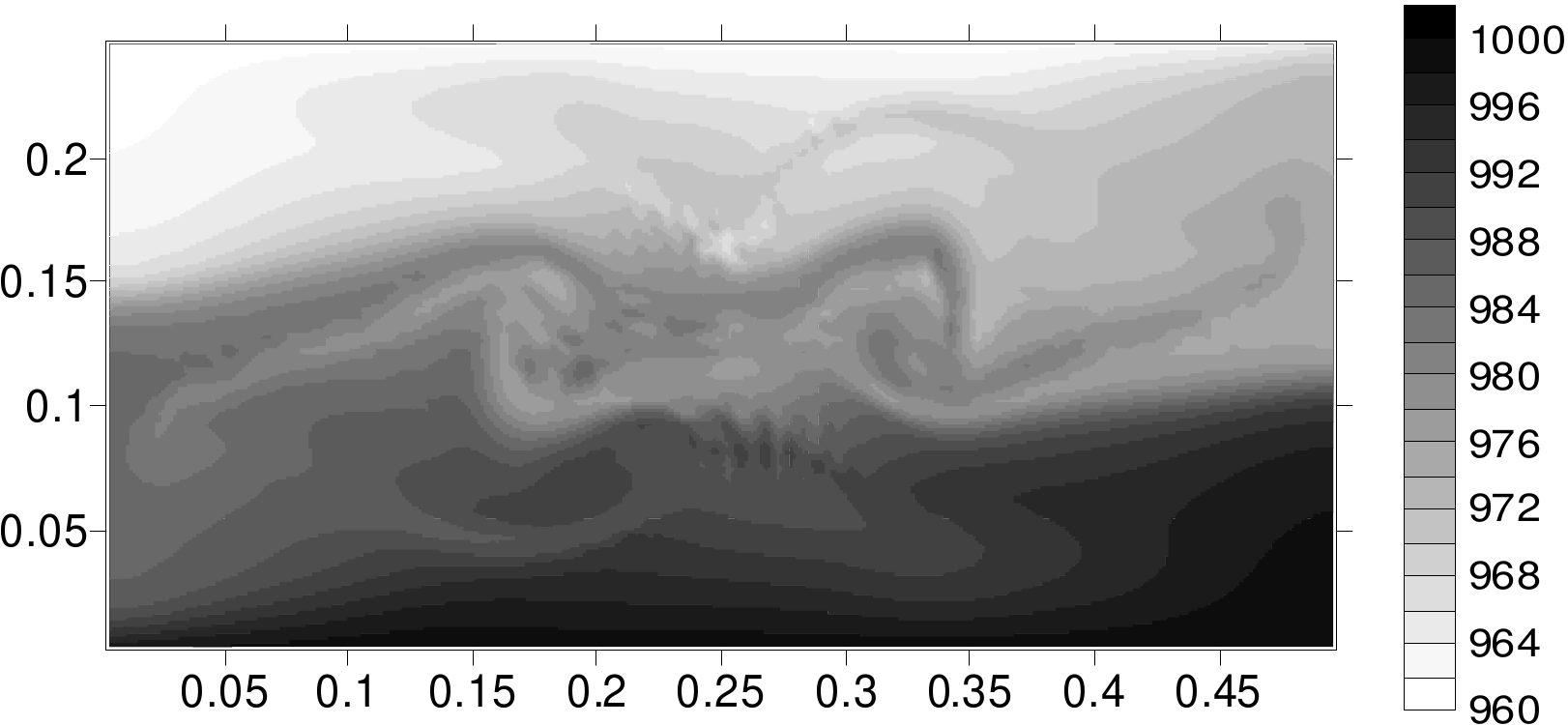}
\caption {Fluid  density  at $t=7 \, \text {s} $. $H=6.23 \, \text
{m} $,  $h=0.25 \, \text {cm} $.} \label{r1-38}
\end{figure}

\begin{figure} [htbp]
\includegraphics [width=12.1166cm] {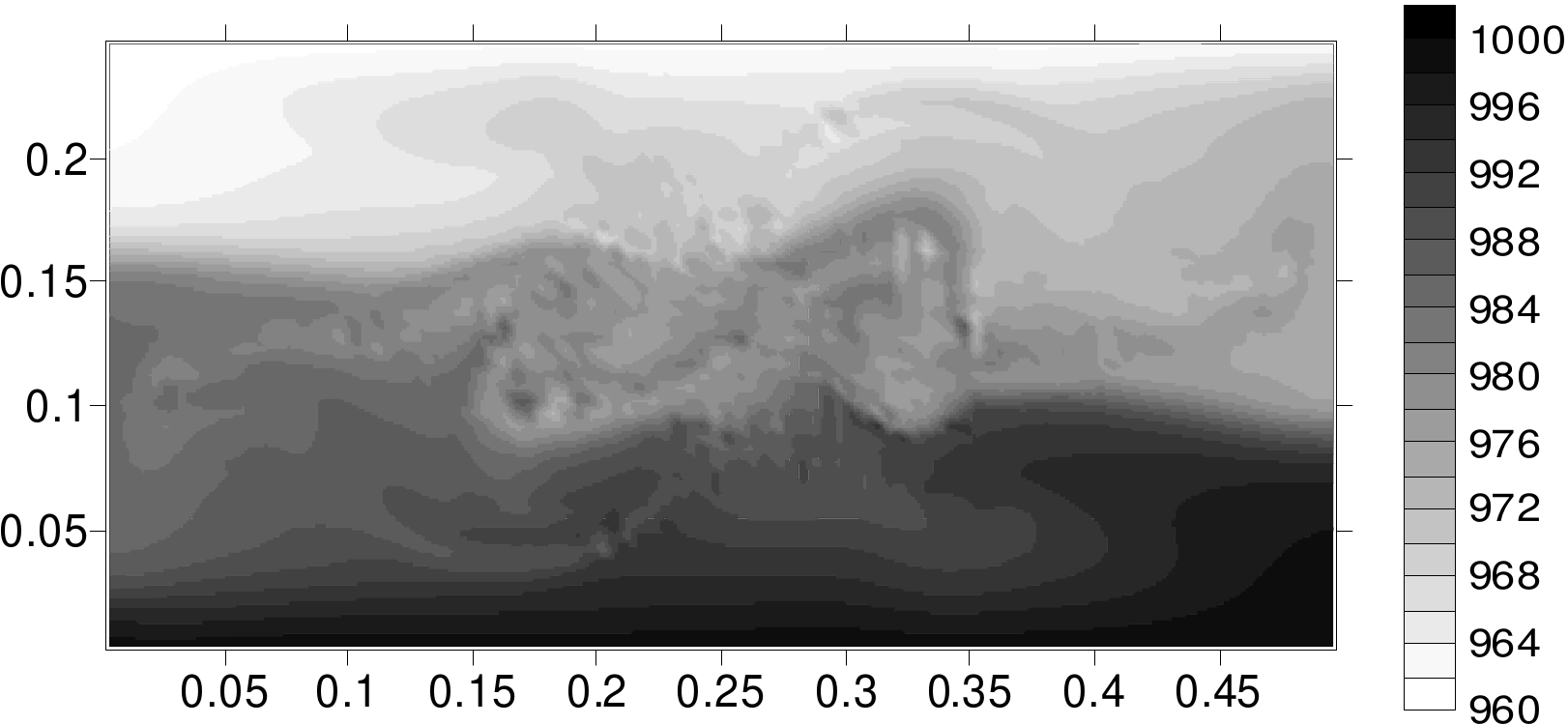}
\caption { Fluid  density  at $t=8 \, \text {s} $. $H=6.23 \,
\text {m} $,  $h=0.25 \, \text {cm} $} \label{r1-48}
\end{figure}

The solution procedure was grounded on numerical integration of the system
 of Euler equations (\ref{Eulerequation1}) by finite-difference method.
 The fluid density for  $t=3\, \text {s} $,
 $7 \, \text {s} $, $8 \, \text {s} $ is shown in fig. \ref{r1-18}, \ref{r1-38}, \ref{r1-48}.


  The overturning and formation of small-scale structures is advancing by formation
  of a tongue of a heavy liquid, penetrating into strata of a light liquid.
  Some layer structure containing segments
  with inverse density is generated as a result.
  At $t=5 \, \text{s}$, abruptions arise in tongues, and isolated fragments
  of a light liquid arise inside a heavy liquid.  At
  $t=7  \, \text{s}$, the process of breaking becomes intensive.  At $t=9 \, \text{s} $,
  about $10 \% $ of fluid in the tank is retracted in intensive small-scale convection.
  The vortex is atomized and the square of the convective spot slowly increases.
  However, with the course of time, small-sized blobs of a heavy liquid subside downwards,
  and blobs of a light fluid go upward.  As a result, stable stratification is
  practically restored to the moment $t=14  \, \text{s}$.


As a whole, the picture of wave breaking qualitatively coincides
with the one circumscribed in \cite{McEwan1971}, \cite{McEwan1}.
However, the last stage of reestablishing of continuous
stratification is absent.  It is explained by ignoring dissipative effects in the model.

\begin {figure} [htbp]
\includegraphics [width=12.1166cm] {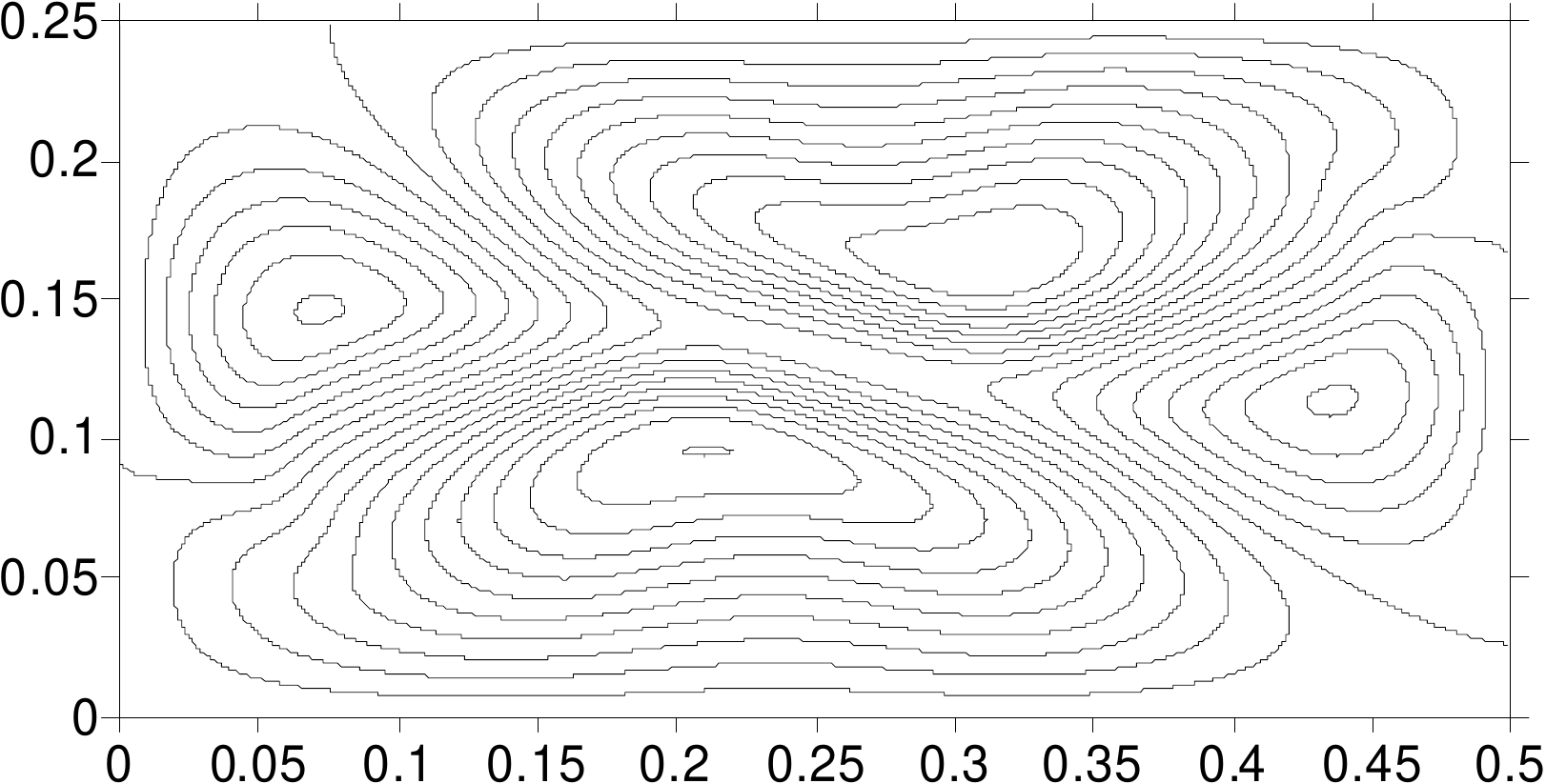}
\caption {Lines of the flow function at $t=3 \, \text {s} $.
$H=6.23 \, \text {m} $,  $h=0.25 \, \text {cm} $} \label{p1-18}
\end {figure}

\begin {figure} [htbp]
\includegraphics [width=12.1166cm] {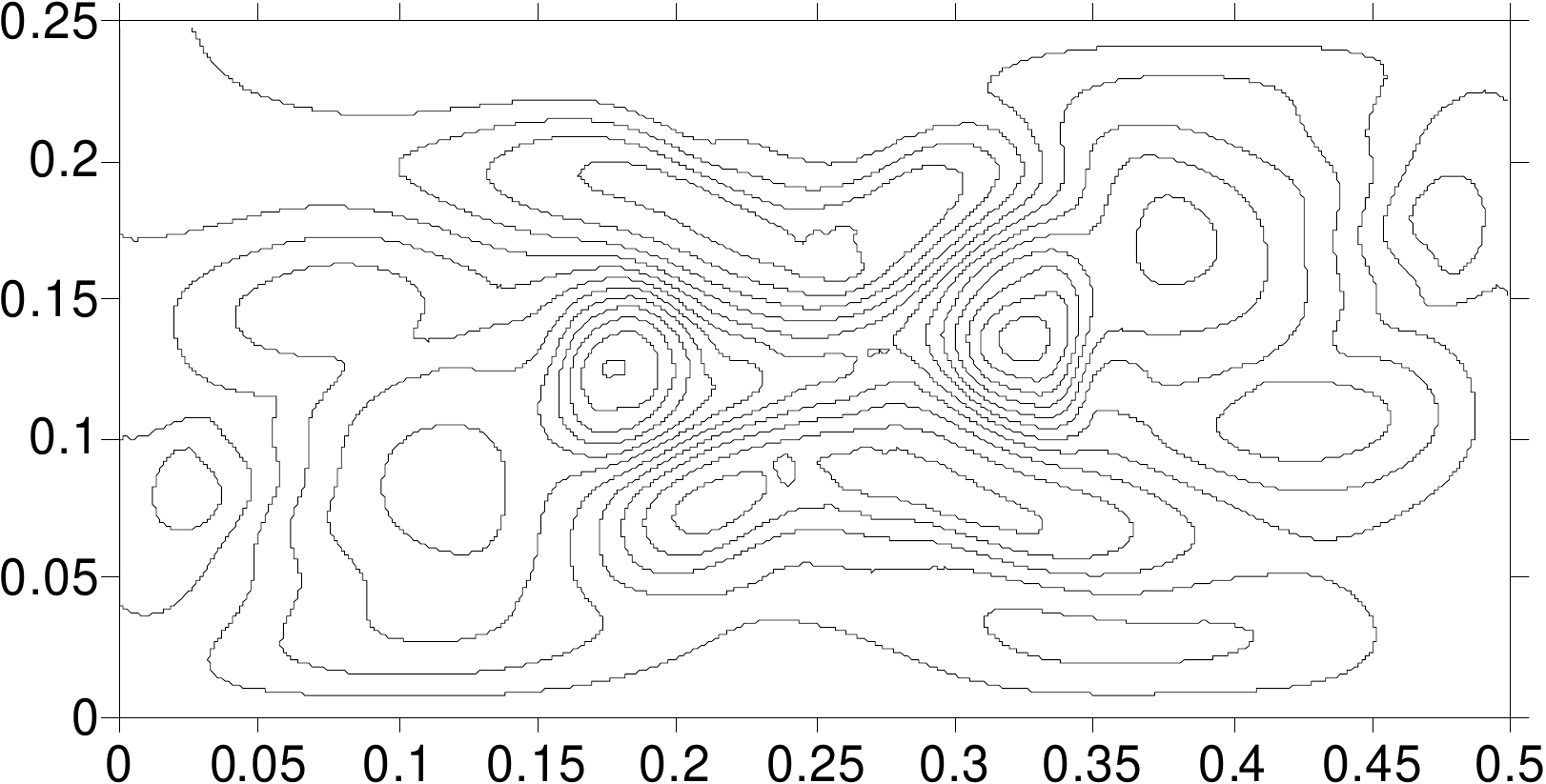}
\caption {Lines of the flow function at $t=7 \, \text {s} $.
$H=6.23 \, \text {m} $,  $h=0.25 \, \text {cm} $} \label{p1-38}
\end {figure}

The flow function is shown in figs. \ref{p1-18}, \ref{p1-38}.
 At the first stage, the disturbance behavior is
typical for internal waves and is perfectly explained by the
theory of small-amplitude internal waves:  left-hand and
right-hand waves arise from this vortex.  Because of the initial
density perturbation is equal zero, the field of density
perturbation is antisymmetric, and the field of a flow function is
centrally symmetric. The symmetry is maintained in good
approximation even when  irregular movement is developed.

Irregular structures in the flow function appear later than in
the density. Perhaps, it is explained by the fact that a flow
function is an integral of velocity, hence it is  the most smooth of
considered physical fields.  Two opposite jet flows are formed at
$t=3 \, \text { s} $, and then they in unison create considerable
velocity shift. The energy of small-scale waves is scooped from a considerable
kinetic energy of the vortex due to
instability development.
 It explains great intensity of the
developed small-scale convection.

\begin {figure} [htbp]
\includegraphics [width=12.1166cm] {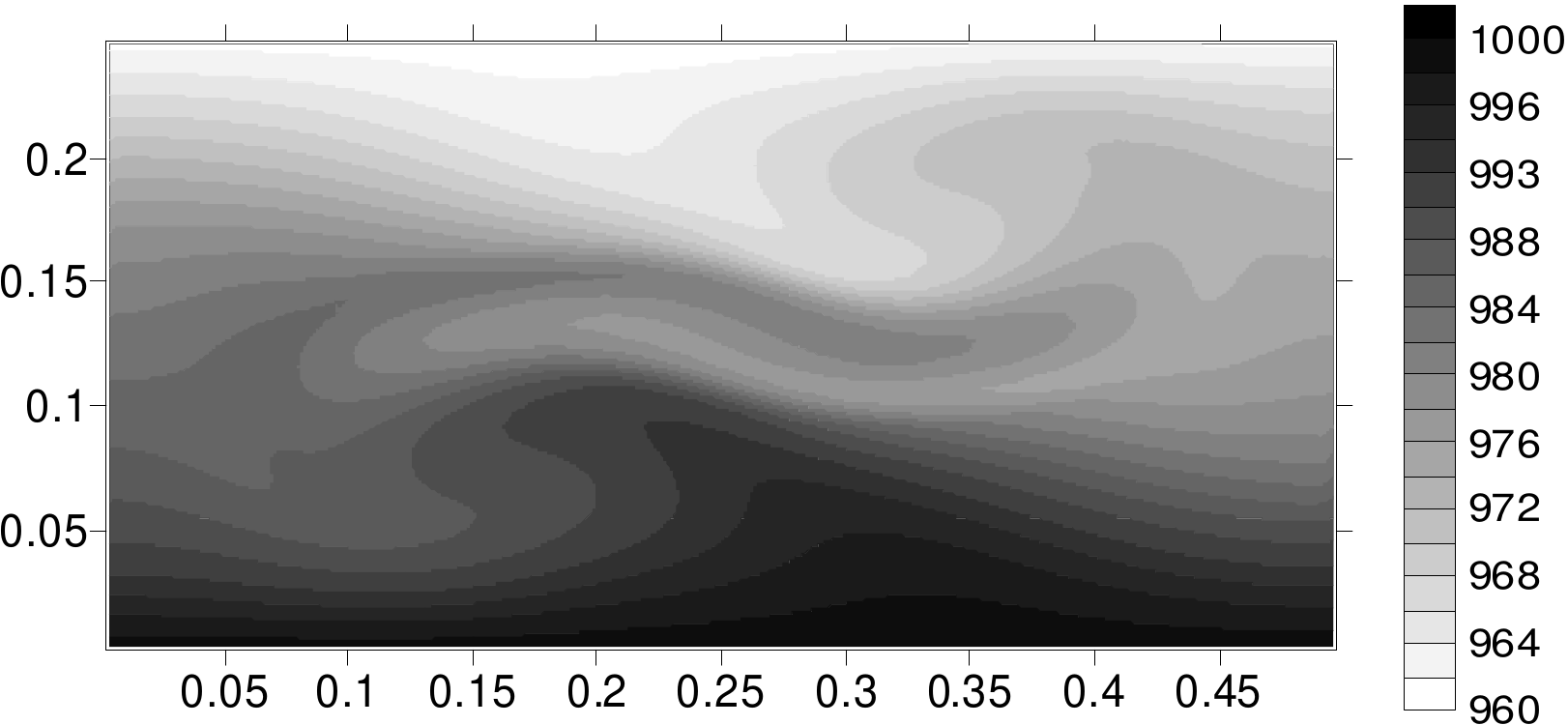}
\caption {Fluid density  at $t=3 \, \text {s} $. $H=6.23 \, \text
{m} $, $h=0.5 \, \text {cm} $} \label{r1-18-1}
\end {figure}

\begin {figure} [htbp]
\includegraphics [width=12.1166cm] {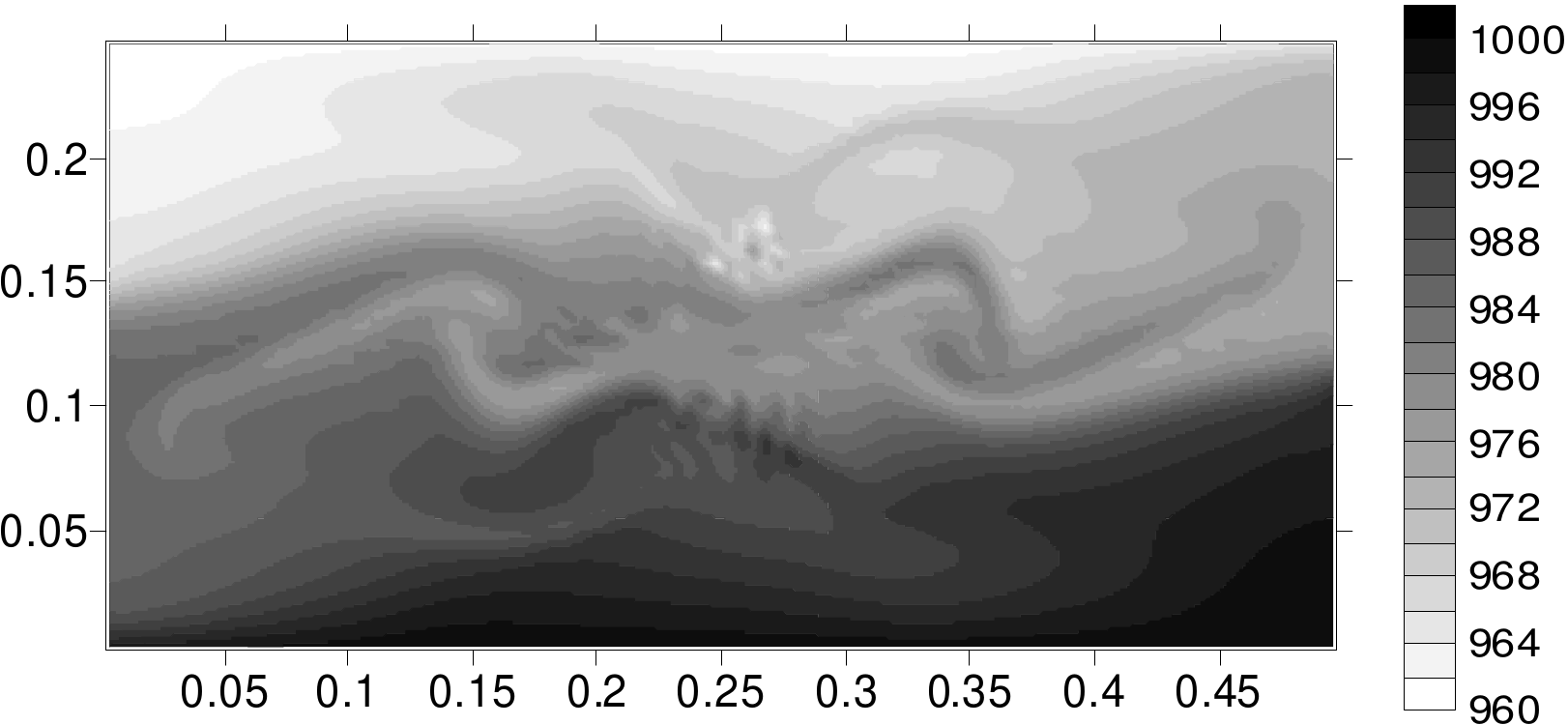}
\caption {Fluid density at $t=7 \, \text {s} $. $H=6.23 \, \text
{m} $, $h=0.5 \, \text {cm} $} \label{r1-38-1}
\end {figure}

To verify exactness of our simulation, the outcomes of simulation
with larger spatial step  $h=0.5 \, \text{cm } $ are shown for
comparison in fig. \ref{r1-18-1},  \ref{r1-38-1}.

At  $t\lessapprox 3 \,\text {s} $, the simulation with the grid
step $h=0.5 \,\text {cm} $ coincide with the simulation with the
scale $h=0.25 \,\text {cm} $. At $t=5 \,\text { s} $, some
small-scale structures arise, and simulations with different steps
differ a little.  The distinctions increase with time.  However,
rather high concurrence of the outcomes is kept. All large-scale
details coincide.

\subsection {Study  of dependence of a vortex destruction on a stratification scale}

\begin {figure} [htbp]
\includegraphics [width=12.1166cm] {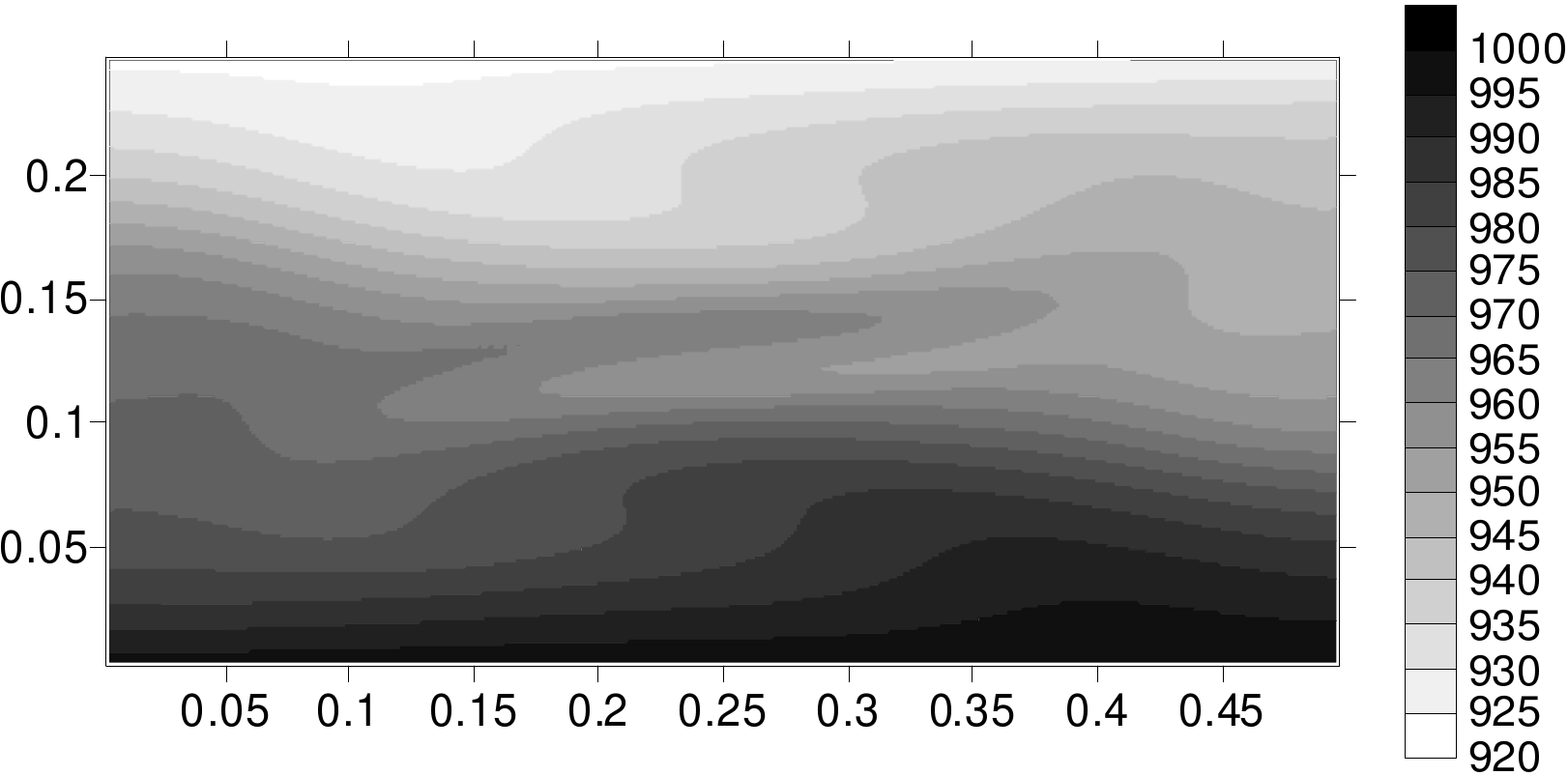}
\caption {Fluid density in the tank at $t=3 \, \text{s} $. $H=3.1
\, \text{m}$,  $h=0.25 \, \text{cm}$} \label{r1-18-d2}
\end {figure}

\begin {figure} [htbp]
\includegraphics [width=12.1166cm] {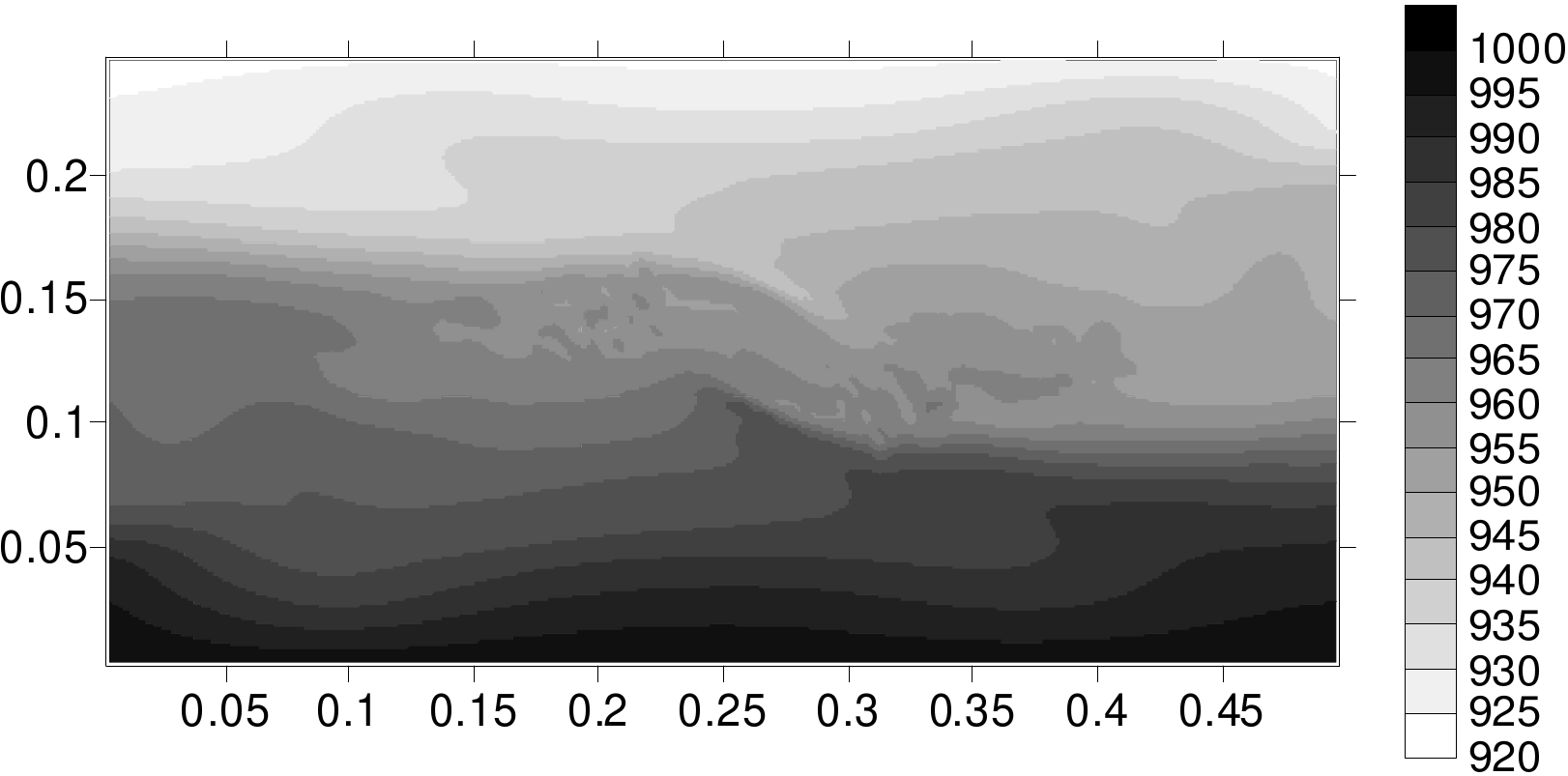}
\caption {Fluid density in the tank at $t=7 \, \text{s} $.
$H=3.1\, \text{m} $,  $h=0.25 \, \text{cm} $} \label{r1-38-d2}
\end {figure}

\begin {figure} [htbp]
\includegraphics [width=12.1166cm] {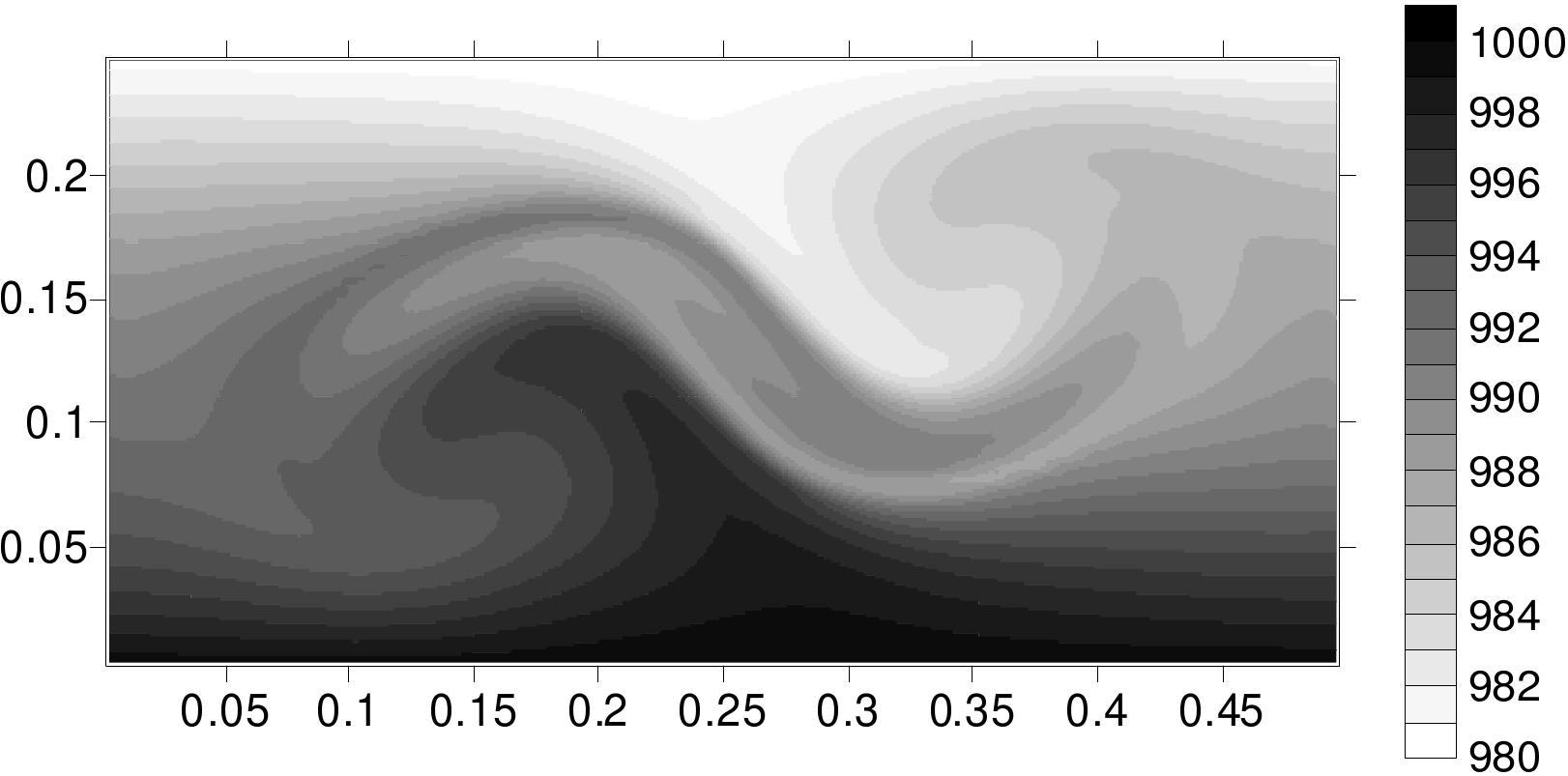}
\caption { Field density in the tank at $t=3 \, \text{s} $.
$H=12.4  \, \text{m}$,  $h=0.25 \, \text{cm}$} \label{r1-18-m2}
\end {figure}

\begin {figure} [htbp]
\includegraphics [width=12.1166cm] {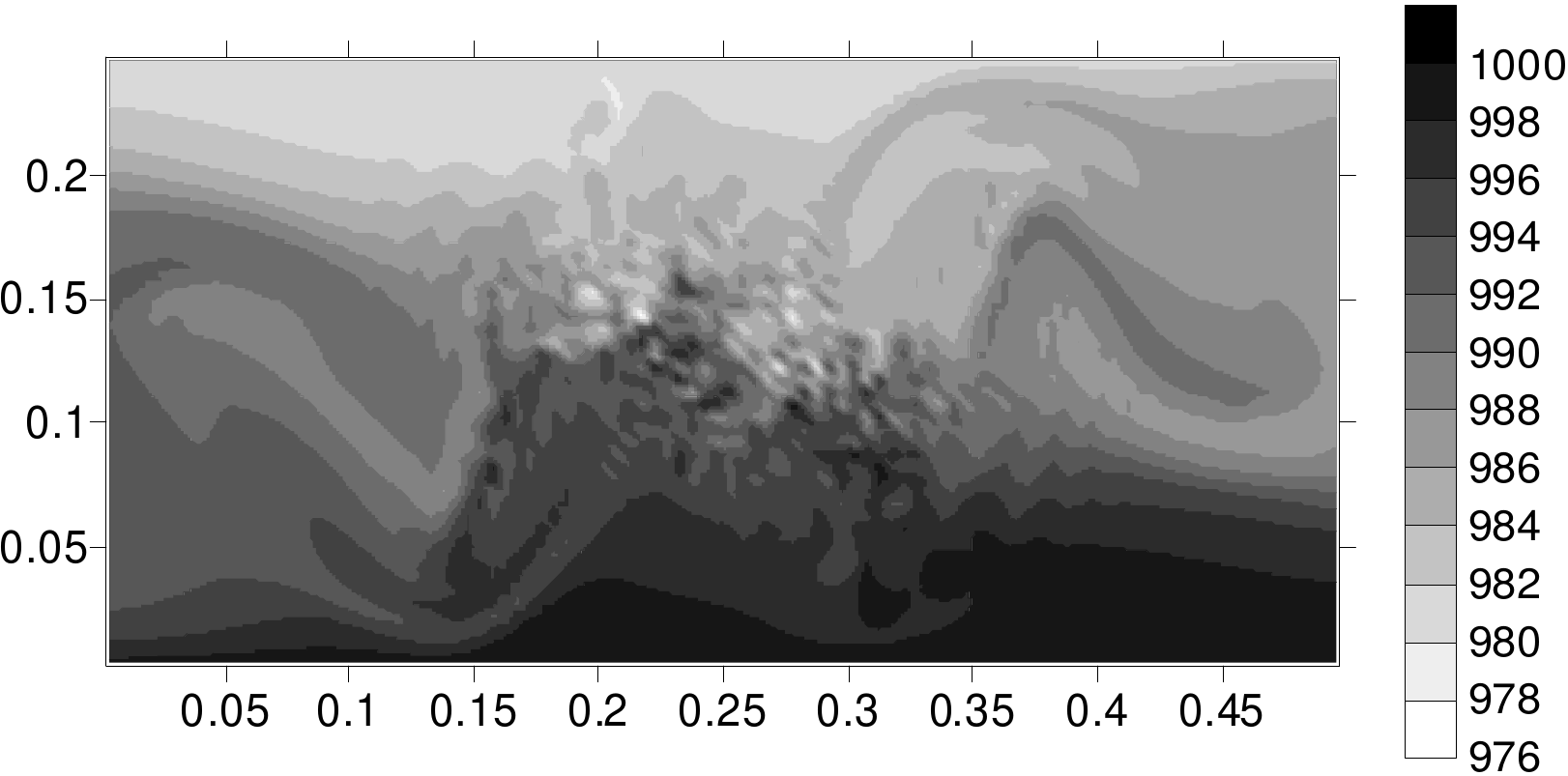}
\caption {Fluid density in the tank at $t=7 \, \text{s} $. $H=12.4
\, \text{m}$,  $h=0.25 \, \text{cm}$} \label{r1-38-m2}
\end {figure}

Identical simulations of a vortex destruction for stratifications  with doubled and twice diminished
 $H $ have been carried out for study  of dependence of the
phenomena on a stratification scale. Some outcomes for $H=3.1 \, \text
{m} $ are shown in fig.  \ref{r1-18-d2}, \ref{r1-38-d2}. We see
that when we decrease  $H $, the evolution becomes an usual  wave process. Fluid
movement becomes more horizontal.  The effect of wave breaking is
starting later in spite of the fact that the inverse  Brunt--V\"{a}is\"{a}l\"{a}
frequency is less.  The wave breaking develops more slowly, and
destruction runs languidly.  Regular wave motion of the fluid is
maintained as a whole, and only separate "winged nuts" \, are
generated.
In fig. \ref{r1-18-m2}, \ref{r1-38-m2}, some outcomes of simulation of
the same wave, but for stratification  with  doubled value of $H $, are
shown.  Comparison between the simulations relating to normal value of $H
$ and the simulations relating to diminished twice $H $ displays
that the phenomenon of vortex destruction and generation
of small-scale convection is increased as  $H $.

\begin {figure} [htbp]
\includegraphics [width=12.1166cm] {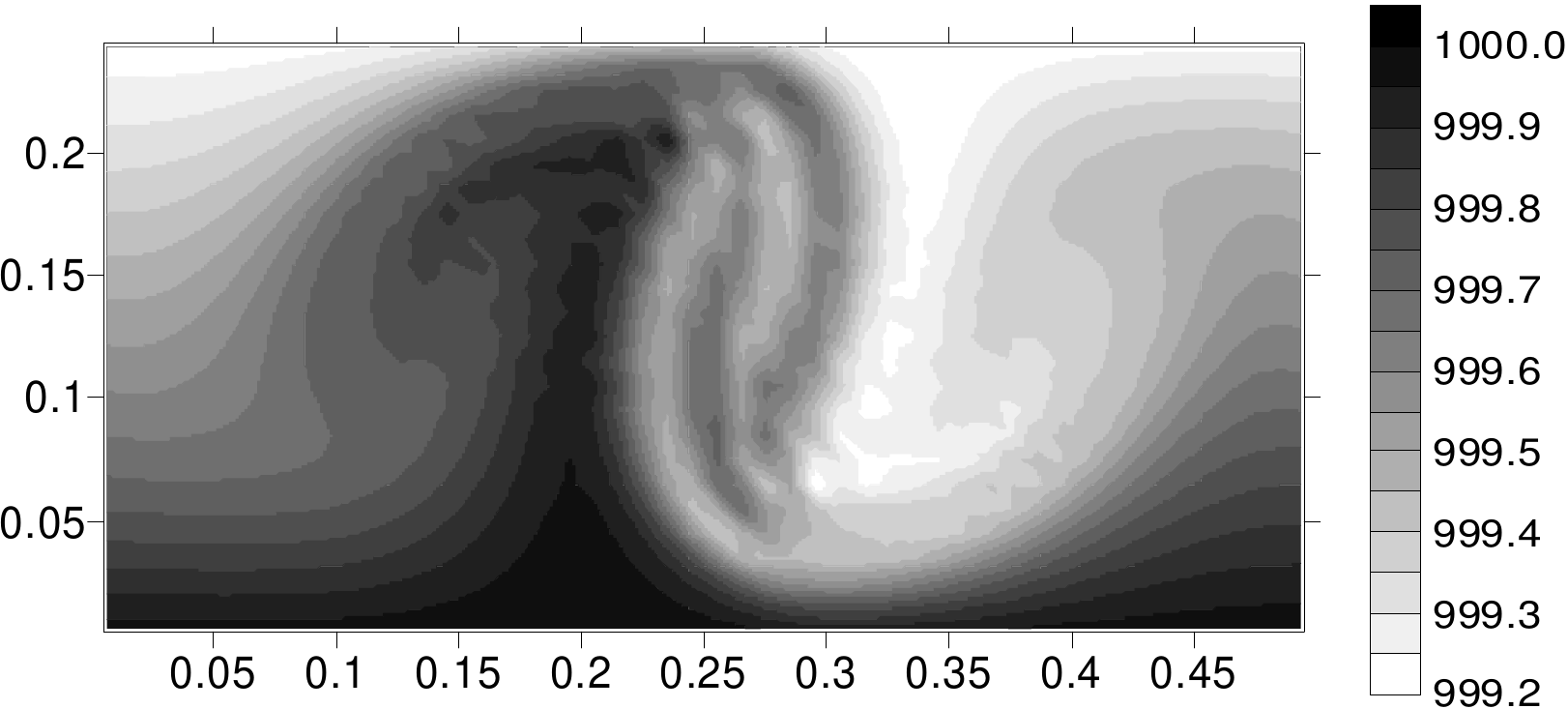}
\caption {Fluid  density in the tank at $t=3 \; $ s. $H=311.5 $ m.
$h=0.25 $ cm.} \label{r1-18-50}
\end {figure}

\begin {figure} [htbp]
\includegraphics [width=12.1166cm] {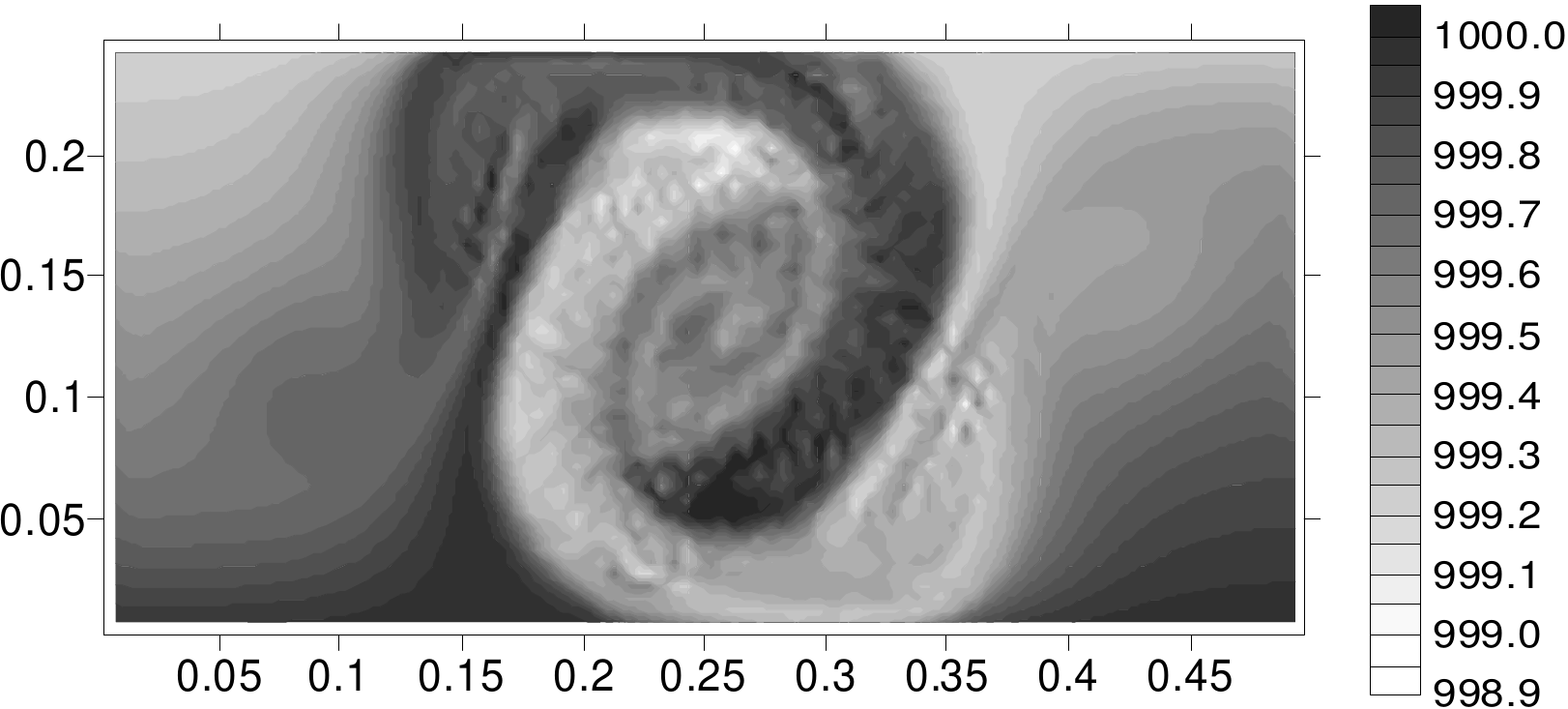}
\caption {Fluid density in the tank at $t=6 \;\text {s} $.
$H=311.5 $ m.  $h=0.25 $ cm.} \label{r1-28-50}
\end {figure}

\begin {figure} [htbp]
\includegraphics [width=12.1166cm] {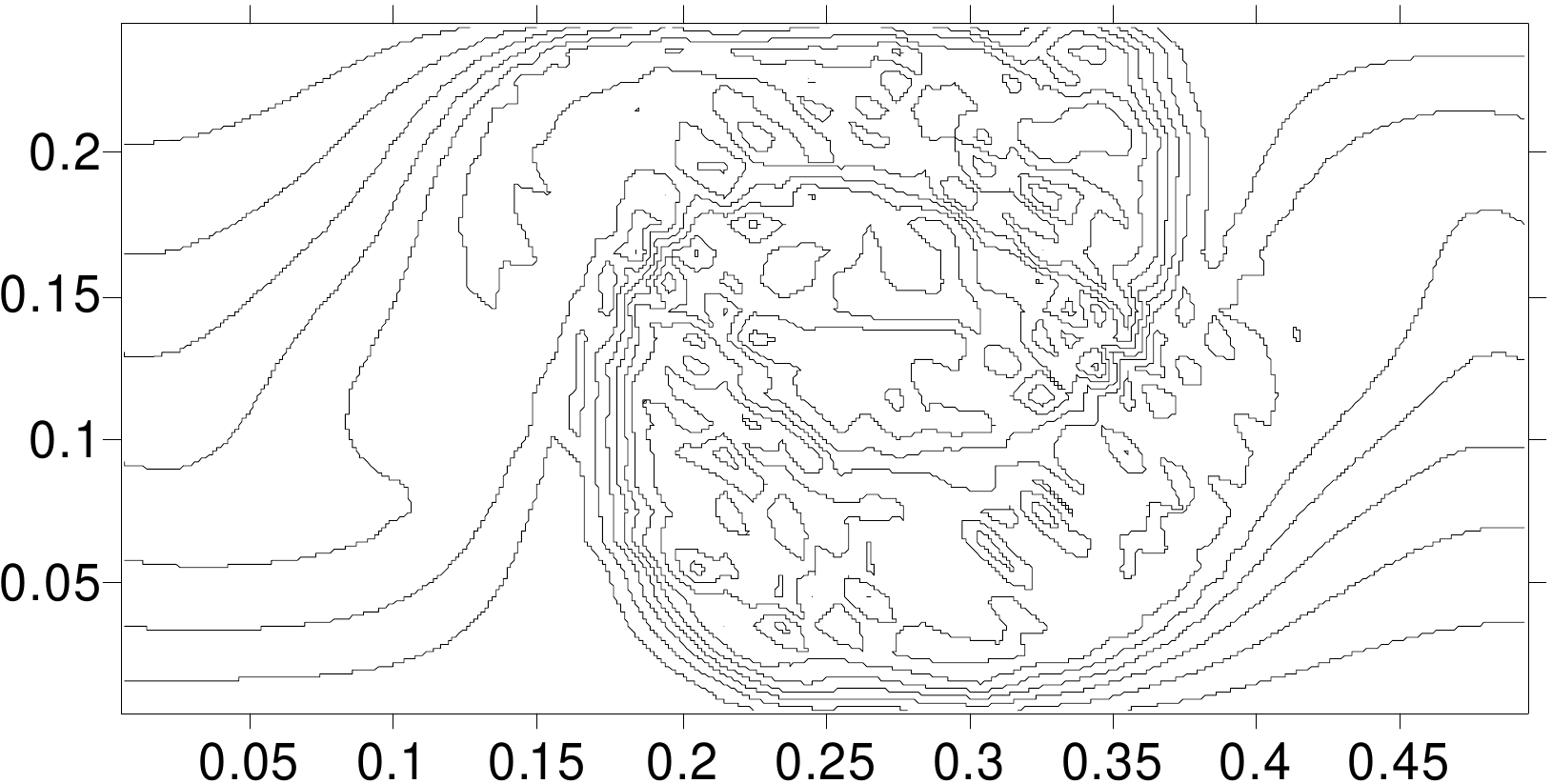}
\caption {The flow function lines at $t=7 \; \text{s}$ in a
stratication with  $H=311.5 $ m. $h=0.25 $ cm.} \label{p1-28-50}
\end {figure}

Therefore, we see that the vortex destruction effect grows with
$H$, and a case of almost homogeneous fluid may be of interest.
To consider the almost homogeneous fluid,  we have carried out test simulation of evolution of the
starting vortex for a stratification with very large $H=311,5 \, \text {m}
$.  The considered fluid is almost homogeneous and the density is
varied only by 0.08 \%.  The simulation outcomes are shown in Fig.
\ref{r1-18-50}, \ref{r1-28-50}, \ref{p1-28-50}.  We see the wave
collapses, and the time of starting of wave breakdown is the same
as in case $H=12.4 $ m. Therefore, for large $H $, the time of
wave breakdown is determined not so much by value of $H
$, as   by starting conditions. If we consider the flow
function, we discover the effect of formation of oppositely directed jet flows with
velocity shear between them.
Due to appearing  this velocity shear, the
instability arises.  Small-scale fluctuations appear in the
density field, and with some retardation, they appear in the flow
function.

\begin{figure}[htbp]
\includegraphics [width=12.1166cm] {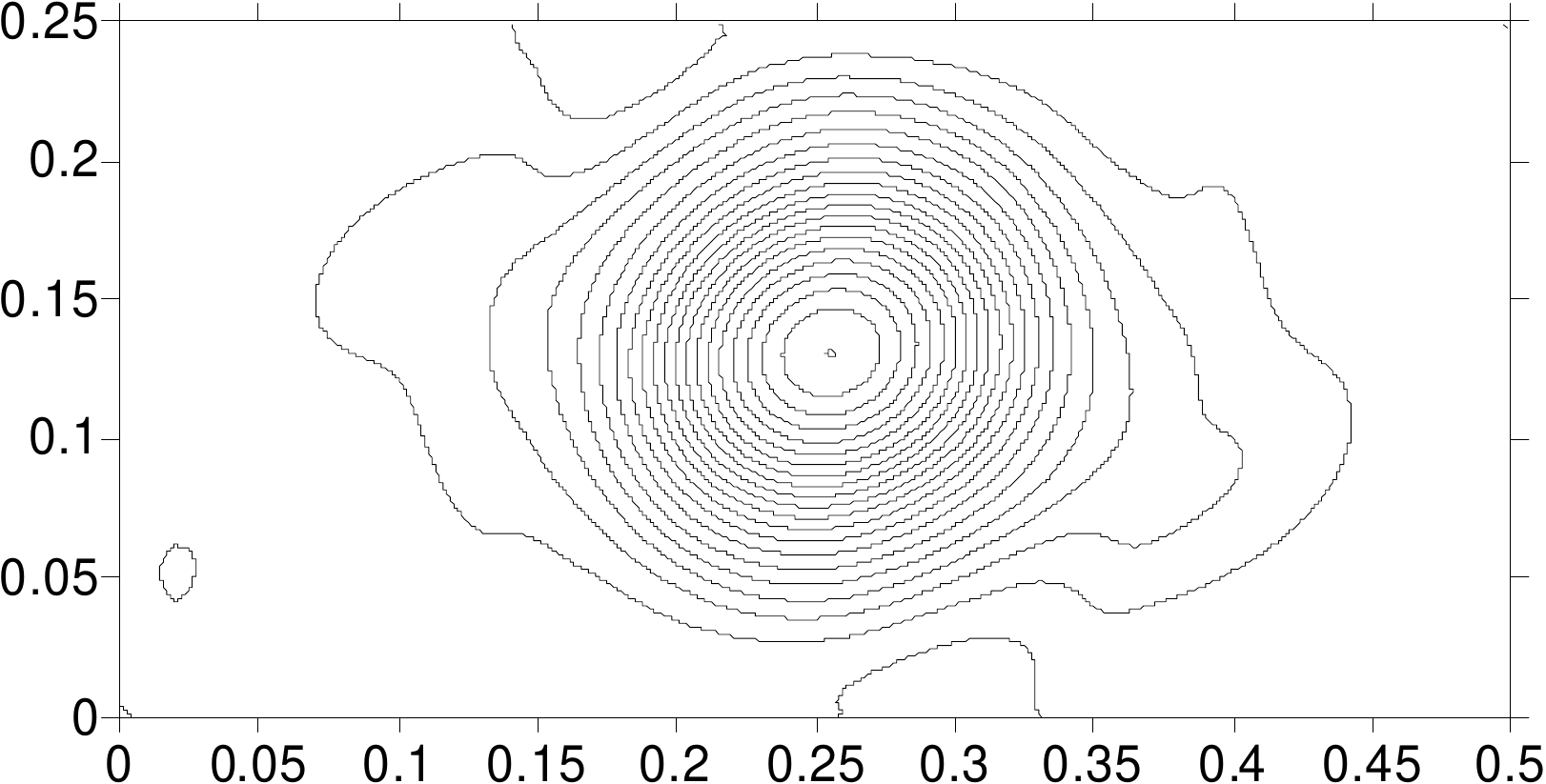}
\caption {The flow function lines at $t=7 \; \text{s}$ in a
homogeneous fluid.} \label{p1-43-hom}
\end{figure}

Some late instant of evolution of the flow function for the same
initial vortex, but in the fluid of a constant density, is shown
in fig. \ref{p1-43-hom}. The fluid density is  constant and is
equal to the typical water density.  We see that the vortex
breakdown and formation of small-scale convection is absent in a
strictly homogeneous medium, and only the vortex has become more symmetric.
Perhaps, it is explained by the fact that any function  $\psi (R)$, $R=\sqrt(x^2+z^2)$ gives stationary solution to equations for homogeneous fluid. Therefore, the vortex takes a symmetric form in time.

Numerical simulations reveal that the
phenomenon of breakdown becomes more and more brightly with
increase of stratification scale $H $.  Nevertheless, the
phenomenon of vortex destruction is absent, when the fluid density
is strictly a constant.  It points out that continuous limit from
a stratified fluid to the case of a fluid of  constant density
is absent for our initial conditions.  The approximation of
constant density is out of the physical sense in such cases, even
if  density variations are very small.

\section {Explanation of the effect}

Below we try to explain why evolution of a vortex in  fluid of
strict-constant density qualitatively differs from evolution of a
vortex in  fluid whose density is almost constant and  varies within
very narrow range.

Let us take  $ \rho\equiv 1 $ for simplicity.  Then the set of equations follows
from (\ref{Eulerequation1})
 \begin{eqnarray}
\frac {\partial \xi} {\partial t} + u \, \frac {\partial \xi} {
\partial x} + w \, \frac {\partial \xi} {
\partial z} =0, \label{equationxi} \\
\frac {\partial^2 \psi} {\partial x^2} + \frac {\partial^2 \psi }
{
\partial z^2} = \xi.  \label{equationxi_1}
 \end{eqnarray}
 In (\ref{equationxi}), $ \xi = \frac {\partial u} {\partial z} -
\frac {\partial w} {\partial x} $, $u (x, z, t) = \frac {\partial
\psi } {\partial z} $, $w (x, z, t) =-\frac {\partial \psi}
{\partial x} $. We assume that $ \xi (x, z, 0) $ is a continuous
function, and $ \max (| \xi (x, z, 0) |) \leq M $.  At the first step, we formally write a solution to (\ref{equationxi}) using a characteristic method and  supposing the functions   $u (x, z, t) $, $w (x, z, t)  $ being known.
 The characteristic method gives us that  $ | \xi (x, z, t) | < sup   (  | \xi (x, z, 0) |  ) $ and $ \xi (x, z, t) \in
L_p (\Omega) $ for any $p> 0 $. Applying the corollary of Sobolev
embedding theorem \cite{Sobolev}, \cite{Ladyzenskaya-1973}, we
arrive at conclusion that the function $\psi (x, z, t) $ is
continuous
 with respect to spatial variables, and  $u (x, z, t) $, $w (x, z, t) $
 are bounded ones.  Moreover,  the characteristic method also shows that  all
evolution of the starting vortex is   only vortex deformation, but
vortex destruction is impossible.

In case of weakly stratified fluid, the  equation for $ \xi $ is like this
 \begin{eqnarray}
 \frac {\partial \xi} {\partial t} + u \frac {
\partial \xi} {
\partial x} + w \frac {\partial \xi} {
\partial z} =Q, \label{equationxi1} \\
\rho \,Q = g \frac {\partial \rho} {\partial x}
 - \frac {\partial \rho} {\partial
z} \left [\frac {\partial u} {\partial t} + u \frac {\partial u}
{\partial x} + w \frac {\partial u} {\partial z} \right] + \frac
{\partial \rho} {\partial x} \left [\frac {\partial w} {\partial
t} + u \frac {\partial w} {\partial x} + w \frac {\partial w}
{\partial z} \right] . \notag
 \end{eqnarray}
The equation for $ \psi $ is the same, as (\ref{equationxi_1}).
The right-hand terms   containing square brackets are small
against the first right-hand term in $Q$.  Therefore, we ignore
them  and we consider only the approximation  $\rho \,Q
\approx g \frac {\partial \rho} {\partial x}$.  We discussed  at
beginning of the paper that a starting vortex in a stratified
fluid by all means shapes tongues of light and heavy fluids  which
 become very thin in   time. Therefore,  very thin density structures
arise  in  time. So, after some evolution, the term  $|g \frac {\partial \rho} {\partial x}|$
becomes  essential.
This  source $Q$  is of small scales and the one  generates new vortices of
smaller scales. These new small-scale vortices arise  in weakly
stratified fluid, and they should generate vortices of smaller scales in due time by means of the mechanism
they arisen themselves.
Therefore, we have  a cascade process of generation of vortices of more and more small scales.

\begin{example}
Any function $\psi \left(R \right) $, where  $R=\sqrt { x^2+z^2  }$ and $\psi \left( R \right)$ vanishes at
infinity, satisfies  (\ref{equationxi},\ref{equationxi_1}). It  gives a stationary solution.
For example, we can take
\begin{equation}
\psi \left( R \right)  = -  \sqrt { \frac {g}{ L}}  \left( R+L\right) \exp \left( - \frac{R} { L} \right)  \, ,
\label{solution}
\end{equation}
where $L$ is a  positive constant.  We can easily do some analytical evaluations this such a function $\psi$. This current function describes  rotation of the fluid with velocity $v \left( R\right) =   \frac {d \psi }{d R}= \sqrt { \frac {g}{ L}}  \frac {R}{L} \exp \left( -R/L \right) $ dependent on the distance from the vortex center. This velocity field has maximum at $R=L$. We can consider (\ref{solution}) as approximate solution to (\ref{equationxi1}), when we neglect of the term $\rho \,Q
\approx g \frac {\partial \rho} {\partial x}$. Then we  substitute (\ref{solution}) in continuity equation and  calculate
\begin{equation}
\rho \left( x,z,t \right) =\rho _{0} \left( z- R* \sin \left( \frac   {v(R) t} {2 \pi R} \right)    \right) =
 \rho _{0} \left( z- R* \sin \left(  \sqrt { \frac {g}{ L}}  \frac   {\exp \left( -R/L \right) t} {2 \pi L} \right)    \right).
\end{equation}
At the following step, we calculate the source
\begin{eqnarray}
\rho \,Q
\approx g \frac {\partial \rho} {\partial x}= g  \rho _{0} ^{\prime } \left( z- R* \sin \left(  \sqrt { \frac {g}{ L}}  \frac   {\exp \left( -R/L \right) t} {2 \pi L} \right)    \right) * \\
* \left[ -\frac{x} {R}* \sin \left(  \sqrt { \frac {g}{ L}}  \frac   {\exp \left( -R/L \right) t} {2 \pi L} \right)    +  \cos   \left(  \sqrt { \frac {g}{ L}}  \frac   {exp \left( -R/L \right) t} {2 \pi L} \right)  *
\left( \sqrt {  \frac {g}{ L} } \frac   {\exp \left( -R/L \right) t x} {2 \pi L}  \right)
 \right]   \notag
\end{eqnarray}
Let stratification be linear and $\frac  {d \rho_o} {dz } =-a $. We see that spacial derivatives of $\rho$ increase in time up to infinite values, and it is the reason we have used a technique  of generalized solutions. For small $t$ we can neglect of the terms $ u \frac {
\partial \xi} {
\partial x} + w \frac {\partial \xi} {
\partial z} $ in ( \ref{equationxi1}) and   integrate the  equation ( \ref{equationxi1}):
\begin{eqnarray}
\xi=\xi_0+\xi_1  \notag   \\
  \xi_0= -  \sqrt { \frac {g}{ L} }  \left( 2 - \frac { R} {L} \right)  \exp \left( - \frac{R} { L} \right)  \notag \\
\xi_1=-\frac {agx}{RL}  \left\lbrace  2 \pi \left( R+L \right)    \left[  \cos  \left( \frac {\exp \left( -\frac {R}{L}  \right) t } {2 \pi}  \right) -1 \right]
 \exp \left( \frac {R}{L}  \right) +R t  \sin   \left( \frac {\exp \left( -\frac {R}{L}  \right) t } {2 \pi}  \right)
 \right\rbrace     \label{xi}
 \end{eqnarray}
 Here $\xi_0$ corresponds to (\ref{solution}), and $\xi_1$ describes generation of new vortices. The expression for   $\xi_1$ contains
 growing term $ -\frac {agx t}{L}   \sin   \left( \frac {\exp \left( -\frac {R}{L}  \right) t } {2 \pi}  \right) $, and due to presence of $ \sin   \left( \frac {\exp \left( -\frac {R}{L}  \right) t } {2 \pi}  \right)$ this term starting some t is a function changing sign as a function of coordinates. It displays generation of new vortices in time owing to stratification influence.
\end{example}

We also note that other  right-hand terms in $Q$  in (\ref{equationxi1}) take into
account  inertial forces arising in flow moving with acceleration.
This inertia forces are equivalent to presence of some gravity field.
That is, the effect of cascade vortex destruction in a nonuniform
fluid can  start without gravity field, but when we apply
accelerated movement  to vortex motion.

\section { Basic outcomes and conclusion}

The system of Euler equations for an incompressible stratified
fluid was studied.  A non-negative nonincreasing functional
extending wave energy functional of the theory of linearized Euler
equations is suggested.  The functional value is conserved over
differentiable solutions, and diminishes if density
discontinuities are present.  Functional properties have allowed
using it for analysis of statement of a generalized problem.  It
is shown that Euler equations are not enough for statement of a
correct generalized problem in case of a stratified fluid in the
gravity field.  Some auxiliary conditions are formulated and
justified.  One of them is an energy conservation law, and the
second is a special requirement for density (\ref {equationforf}).
The statement of a generalized problem is formulated.  The
finite-difference problem-solving procedure is developed.
Existence of a weak solution is proved.

The phenomenon of destruction of starting vortex in conditions of
McEwan's laboratory experiments \cite{McEwan1971}, \cite{McEwan1}
has been simulated and studied.  Qualitative concurrence of the
simulated effect with the one observed in laboratory experiments,
is good, but the last stage of restoring of smooth stratification
is absent due to usage of an ideal liquid model.  Some new data
about the phenomenon, difficult for laboratory measurement, are
obtained, and evolution of the flow function is studied.  It is
shown that large gradients appear in the velocity field at the
expense of nonlinear effects and they may play an important role
in development of  instability.

By means of numerical experiments, dependence of the solution on a
stratification scale $H $ is studied.  It is revealed that the
effects of vortex destruction and formation of   small-scale
convection grows with $H $ for strongly nonlinear waves.  On the
contrary, when we decrease  $H $, the effect of wave destruction becomes weak.
The wave breakdown is starting from the density field and
some later formation of small-scale structures in the flow
function is observed.  The effect of vortex destruction is not
observed when the fluid density is strictly a constant.

\bibliographystyle{amsplain}

%


\end{document}